\documentclass[runningheads,a4paper,12pt]{article}

\usepackage{cmap}
\usepackage{latexsym}
\usepackage[utf8]{inputenc}
\usepackage[unicode]{hyperref}
\usepackage{fancyhdr}
\usepackage{cmap}
\usepackage{amssymb}
\usepackage{amsmath}
\usepackage{tikz}
\usetikzlibrary{shapes,arrows, positioning}
\pgfmathtruncatemacro\distance{1}

\def\ral#1{\;\mathop{\longrightarrow}\limits^{\!#1}\;}

\def\blackbox{\;\vrule height 7pt width 7pt depth 0pt \;}
\def\enemy{{\dagger}}

\def\be#1{\begin{equation}\label{#1}}
\def\ee{\end{equation}}

\def\re#1{(\ref{#1})}
\def\i{\item}

\def\bi{\begin{itemize}}
\def\ei{\end{itemize}}
\def\bn{\begin{enumerate}}
\def\en{\end{enumerate}}

\def\l#1{\langle #1 \rangle}

\def\c#1{\left\{\begin{array}{lllll}#1\end{array}\right\}}

\def\by{\begin{array}{llllllllllllll}}
\def\ey{\end{array}}

\def\eam{\;\mathbin{{\mathop{=}\limits^{\mbox{\scriptsize def}}}}\;}

\def\ba{\left\{\begin{array}{llllllllll}}
\def\ea{\end{array} \right\}}
\def\bau{\left[\begin{array}{llllllllll}}
\def\eau{\end{array} \right]}
\def\beq{\begin{equation}}

\def\ra#1{\;\mathop{\to}\limits^{
\begin{picture}(0,5)
\put (-3,2){\makebox(0,1){$\scriptstyle #1$}}
\end{picture}
}\;}

\def\blackbox{\;\vrule height 7pt width 7pt depth 0pt \;}

\def\be#1{\begin{equation}\label{#1}}
\def\ee{\end{equation}}
\def\i{\item}

\def\re#1{(\ref{#1})}
\def\bn{\begin{enumerate}}
\def\en{\end{enumerate}}

\def\l#1{\langle #1 \rangle}
\def\c#1{\left\{\begin{array}{lllll}#1\end{array}\right\}}

\def\by{\begin{array}{llllllllllllll}}
\def\ey{\end{array}}
\def\bi{\begin{itemize}}
\def\ei{\end{itemize}}
\def\bn{\begin{enumerate}}
\def\en{\end{enumerate}}
\def\eam{\mathbin{{\mathop{=}\limits^{\mbox{\scriptsize def}}}}}

\def\beenum{\begin{enumerate}}
\def\enenum{\end{enumerate}}
\def\ba{\left\{\begin{array}{llllllllll}}
\def\ea{\end{array} \right\}}
\def\bau{\left[\begin{array}{llllllllll}}
\def\eau{\end{array} \right]}
\def\beq{\begin{equation}}

\def\blackbox{\;\vrule height 7pt width 7pt depth 0pt \;}

\def\be#1{\begin{equation}\label{#1}}
\def\ee{\end{equation}}

\def\re#1{(\ref{#1})}
\def\i{\item}

\def\bi{\begin{itemize}}
\def\ei{\end{itemize}}
\def\bn{\begin{enumerate}}
\def\en{\end{enumerate}}

\def\l#1{\langle #1 \rangle}

\def\c#1{
{\def\arraystretch{0.7}
\left\{\begin{array}{lllll}#1\end{array}\right\}}}

\def\by{\begin{array}{llllllllllllll}}
\def\ey{\end{array}}

\def\eam{\mathbin{{\mathop{=}\limits^{\mbox{\scriptsize def}}}}}

\def\ba{\left\{\begin{array}{llllllllll}}
\def\ea{\end{array} \right\}}
\def\bau{\left[\begin{array}{llllllllll}}
\def\eau{\end{array} \right]}
\def\beq{\begin{equation}}

\def\by{\begin{array}{llllllllllllll}}
\def\ey{\end{array}}

\def\eam{\mathbin{{\mathop{=}\limits^{\mbox{\scriptsize def}}}}}

\def\ba{\left\{\begin{array}{llllllllll}}
\def\ea{\end{array} \right\}}
\def\bau{\left[\begin{array}{llllllllll}}
\def\eau{\end{array} \right]}
\def\beq{\begin{equation}}

\def\Box{\hbox
   {\vbox
     {\hrule height .033em
      \hbox to .5em
        {\vrule width .033em
         \hbox{\vbox to .5em{\vss}\hss}
         \vrule width .033em}
      \hrule height .033em
     }}}

 \newif\ifqed\qedfalse 
 \def\endofpfmarker{\mbox{$\Box$}}
 \def\endofpf{\hbox{\ }\hfill\endofpfmarker\qedtrue}
 \def\proofstyle{}
 \def \qedhere {\endofpf}

 \newenvironment{pfenv}[1]{%
   \global\def\qedhere{\endofpf\global\def\qedhere{}}
   \proofstyle
   \trivlist 
   \item[\hskip \labelsep{\bf #1}]%
   }{%
   \qedhere
   \endtrivlist}

\def\vect#1{\widetilde{#1}}

\makeatletter
 {}
 {}
\makeatother

\newcommand{\bicong}{\approx }

\newcommand{\rel}{\mathrel{\mu}}
\newcommand{\wkbisim}{\mathrel{\approx_l}}

\newcommand{\barb}[1]{\Downarrow{\!#1}}

\def\subst#1#2{^{#1} \!/\! _{#2}}

\newcommand{\sx}[1]{\subst{#1}{x}}

\newcommand{\smx}{\{\sx{e}\}}

\def\ltr#1{\xrightarrow{#1}}
\newcommand{\infrule}[2]{\cfrac{\mbox{$#1$}}{\mbox{$#2$}}}

\newcommand{\Msg}[1]{\langle#1\rangle}
\newcommand{\Abs}[1]{(#1)}
\newcommand{\kw}[1]{\mathit{#1}}

\newcommand{\fv}{\kw{fv}}            

\newcommand{\dom}{\kw{dom}}

\newcommand{\Rcv}[2]{#1\Abs{#2}}
\newcommand{\Snd}[2]{\overline{#1}\Msg{#2}}

\newcommand{\Res}[1]{\nu #1.}
\newcommand{\parop}{\mathbin{\mid}}
\newcommand{\Repl}[1]{\mathord{! #1}}
\newcommand{\nil}{\mathbf{0}}
\newcommand{\IfThen}[3]{\kw{{\bf if}}\ #1 = #2\ \kw{{\bf then}}\ #3}
\newcommand{\IfThenElse}[4]{\kw{{\bf if}}\ #1 = #2\ \kw{{\bf then}}\ #3\ \kw{{\bf else}}\ #4}

\newcommand{\sdecrypt}[3]

\def\tot{\ltr{*}}

\author{Andrew M. Mironov}

\title{
A simple proof of the coincidence of observational and labeled equivalence of processes in applied pi-calculus
}

\date{
Moscow State University
$\;$\\
amironov66@gmail.com
}

\begin{document}

\maketitle 

\newcounter{theorem}
\newcounter{lemma}
\newcounter{fig}

\begin{abstract}
This paper presents a new, significantly simpler proof of one of the main results of applied pi-calculus: the theorem that the concepts of observational and labeled equivalence of extended processes in applied pi-calculus coincide.

\end{abstract}

\section{Introduction}

The applied pi-calculus
was first presented in 2001 in the paper \cite{afpi}.
Its modern presentation can be found in
the paper \cite{apical}.
This calculus was intended for the formal description
and analysis of cryptographic
protocols \cite{crypto}.
Currently, the applied pi-calculus is also used
in problems of modeling and verifying
business processes \cite{business}, in modeling
biomolecular systems \cite{bio},
analyzing multi-agent systems in artificial intelligence,
and in other problems.
One of the main results of applied pi-calculus is the theorem that the concepts of observational and labeled equivalences of extended processes coincide. The proof of this result, along with auxiliary assertions, occupies several dozen pages in the paper \cite{apical}. In this paper, we show that the proof of this result can be significantly simplified.
The author considers slightly modified definitions
of the concept of transitions on extended processes, which do not affect the properties of the extended processes under study.

\section{Syntax of applied pi-calculus}

The main objects of study
in the pi-calculus
are {\bf processes} with synchronous interaction,
for the definition of which we
present the necessary mathematical concepts below.

\subsection {Variables and terms}

We assume that we are given countable
sets of variables $Var$, names $Names$, constants $Con$, and a finite set of functional symbols (FS) $Fun$, where each FS $f\in Fun$ is associated with an arity $ar(f)>0$.
We will denote the set $Var\sqcup Names$ by the symbol ${\cal U}$. The set  $Tm$ of {\bf terms} is defined inductively:
each term $e\in Tm$ is either a variable or name $u\in {\cal U}$,
or a constant $d\in Con$,
or has the form
$f(e_1,\ldots, e_n)$, where
$f\in Fun$, $e_1,\ldots, e_n$ is a list of terms, and
$n=ar(f)$.
We will assume that
$Fun$
contains the FS $pair$, where $ar(pair)= 2$,
and terms of the form $pair(e_1,e_2)$
will be written more concisely as
$(e_1,e_2)$.

$\forall\,e\in Tm$
the notation $var(e)$ denotes the set
of all variables contained in $e$. A term $e$ is said to be {\bf closed} if $var(e)=\emptyset$.
$\forall\,X\subseteq Var$, the notation
$tm(X)$ denotes the set of all terms
$e\in Tm$ that satisfy the condition $var(e)\subseteq X$.
We will denote the lists $x_1,\dots,x_l$, $n_1,\dots,n_l$,
$u_1,\dots,u_l$, and $e_1,\dots,e_l$,
where $x_i\in Var$, $n_i\in Names$,
$u_i\in {\cal U}$, $e_i\in Tm$, $i=1\ldots, l$,
by the notations $\tilde x$, $\tilde n$,
$\vect{u}$, and~$\vect{e}$, respectively. 
For each list of terms $\tilde e$, the notation $var(\tilde e)$ denotes the set $\bigcup_{i=1,\ldots, l}var(e_i)$.
If $E$ is an expression that may contain variables or names (it could be a term, a list of terms, a set of variables or names, or a process expression defined below), then
$\forall\,u\in {\cal U}$
the notation $u\not\in E$ means that $u$ is not in $E$,
$\tilde u\subseteq E$ means that each component of $\tilde u$
is in $E$. If $E$ and $E'$ are expressions of the form specified above, then
$E\not\!\!\cap \,E'$ means that
$E$ and $E'$ do not share any variables.

\subsection{Substitutions}
\label{nsadrew}

A {\bf substitution} is a function
$\theta:X\to Tm$, where $X$ is some set of variables, this set is denoted by
$dom(\theta)$.
The substitution $\theta$
is said to be {\bf acyclic} if $dom(\theta)$
can be represented as a list $\tilde x = x_1,\ldots, x_l$
with the property:
\be{sadfvaew}
\forall\,i,j:1\leq i\leq j\leq l\;\;
x_i\not\in var(\theta(x_j)).\ee

The set of all acyclic substitutions is denoted by $\Theta$.
If
$\theta\in \Theta$ and $dom(\theta)$ is represented as a list
$\tilde x=x_1\ldots x_l$,
then $\theta$ can be written as $\{\subst{\tilde e}{\tilde x}\}$,
where $\tilde e=\theta(x_1)\ldots \theta(x_l)$.
Below, we assume that if
a substitution
is represented as
$\{\subst{\tilde e}{\tilde x}\}$,
then ${\tilde x}$
has the \re{sadfvaew} property.
If $e\in Tm$ and $\{\subst{e'}{x}\}\in
\Theta$, then the notation $e^{\{\subst{e'}{x}\}}$
denotes the term
obtained from $e$
by replacing each occurrence of $x$
in $e$ with $e'$.

If
$\theta=\{\subst{e_1e_2\ldots e_l}{x_1x_2\ldots x_l}\}\in \Theta$,
then
$\forall\,
e\in Tm$
 $e^{\theta}$
denotes the term
$$(\ldots((e^{\{\subst{e_1}{x_1}\}})
^{\{\subst{e_2}{x_2}\}})^{\ldots})^{\{\subst{e_l}{x_l}\}}.$$

\subsection{Processes}

{\bf Processes} of the pi-calculus
are plain processes
and extended processes. Each process is an expression (i.e., a sequence of symbols) constructed from terms and symbols of process operations.

{\bf Plain processes (PP)} are denoted
by the symbols $P, Q, R$ (possibly with subscripts). 
 Below, $c, e, e'$ denote terms, $n$ denotes a name, $x$ denotes a variable, and $P, Q$ denotes PPs.

Each PP has one of the following forms:
\bi\i $\nil$ (null process, does nothing, terminates immediately),
\i $P \parop Q$ (parallel composition, executed by
simultaneously executing $P$ and~$Q$),
\i $\Repl P$ (executes as an infinite number of copies of $P$,
executing in parallel),
\i $\Res n P$ (behaves like $P$, in which
the name $n$ is bound, the concepts of bound names and variables are defined below),
\i $\IfThenElse{e}{e'}{P}{Q}$ (checks the condition $e=e'$; if true,
then executes as $P$, otherwise executes as $Q$),
the notation $\IfThen{e}{e'}{P}$ is shorthand for
$\IfThenElse{e}{e'}{P}{{\bf 0}}$,
\i $\Rcv c x.P$, where $x\not\in c$
(receives a message from channel $c$, and then
executes as $P$ with the received message substituted for
$x$),
\i $\Snd c e.P$
(outputs message $e$ to channel $c$, then executes as~$P$).
\ei

{\bf Extended processes (EPs)} are denoted by the symbols
$A, B, C$ (possibly with subscripts) and are also defined inductively.
Each EP has one of the following forms (below, $P$ is a PP, $A$ and $B$ are EPs, $u\in {\cal U}$):
 $P$ (a plain process),
 $A \parop B$ (a parallel composition of $A$ and $B$),
 $\Res u A$
(behaves like $A$, in which
$u$ is bound),
 $\smx$ (a substitution).

Components of the form $\nu u$ in an EP are called {\bf binding operations}.

\subsection{Bound occurrences of names 
and variables}
\label{fdgfsdhdr5}

An occurrence of
$u\in {\cal U}$ in a EP $A$ is said to be
{\bf bound}
if it is contained
in a subexpression of the form
$\nu u.B$ of expression $A$,
or is a part $(u).P$ of the
subexpression 
$\Rcv c u.P$
of expression $A$.
otherwise,
an occurrence of $u$ in $A$
is said to be {\bf free}.
The set of all variables
or names
that have free occurrences
in $A$ is denoted by $fv(A)$
or $fn(A)$, respectively.

If a process $A$ has the form $\nu u.B$ or $\Rcv c u.P$,
then the {\bf boundness group} of $u$ in $A$
is the set of occurrences of $u$ in $A$,
consisting of the first occurrence of $u$ in $A$ and all occurrences
of $u$ in $B$ or $P$, respectively,
that are free in these processes.
If $A=\nu u.B$ or $\Rcv c u.P$,
then
we consider the process $A$ to be equal to the process
that is obtained from $A$ by replacing all occurrences
of $u$ in the boundness group of $u$ in $A$ with
an arbitrary $u'\in {\cal U}$ that has no occurrences in $A$
(such a replacement is called {\bf renaming} of bound variables
or names that are part of the same boundness group). Below, we assume that in each process under consideration, all bound variables or names are renamed so that they do not appear in any other process under consideration.

\subsection{Closed processes, correct processes}

For each EP $A$, $\dom(A)$ denotes the set
$$\{x\in fv(A)\mid \mbox{$A$ has a substitution of the form $\smx$}\}.$$

A EP $A$ is  a {\bf closed EP (CEP)}
if $\dom(A) = \fv(A)$.
We will denote the set of all CEPs
by  ${\cal P}$.

A EP $A$ is said to be {\bf correct} if
the following properties hold.
\bi
\i If $A$ contains a subexpression of the form
$B|C$, then $$dom(B)\cap dom(C)=\emptyset.$$
\i $\forall\,x\in Var$ $A$ contains at most one substitution of the form
$\{\subst{e}{x}\}$.
\i If $A$ contains a subexpression of the form $\nu x.B$, where $x\in Var$, then $B$ contains exactly
one substitution of the form $\{\subst{e}{x}\}$.
\i The set of all substitutions occurred in $A$ can be
represented as a list
$\{\subst{e_1}{x_1}\}, \ldots, \{\subst{e_l}{x_l}\}$, where
$\{\subst{e_1\ldots e_l}{x_1\ldots x_l}\}$ is acyclic.
\ei

Below, we assume that all the considered EPs are correct.
We will use the following notation: \bi\i $\nu \vect{u}$,
where $\vect{u} = u_1\ldots u_l$, $l\geq 0$, denotes
the (possibly empty) list of the form $\nu u_1.\nu u_2.\dots \nu u_l$,
\i EP
$\{\subst{e_1}{x_1}\} \parop
\dots \parop \{\subst{e_l}{x_l}\}$ is denoted by
$\{\subst{\vect{e}}{\vect{x}}\}$, where $\vect{e}=(e_1\ldots e_l)$,
$\vect{x}=(x_1\ldots x_l)$
(if $l=0$, then this EP is equal to {\bf 0} by definition).
\ei

\subsection{Contexts}

A {\bf context} is 
an expression $E$, possibly containing the symbol $\cdot$
(which is understood as a process variable),
defined inductively: either $E=\cdot$, or $E$ is a EP, or $E$ has the form
$E'|A$ or $\nu u.E'$,
where $E'$ is a context, $A$ is a EP,
and
$u\in {\cal U}$.
If $E$ is a context and $A$ is an EP, then $E[A]$ denotes the result of replacing in $E$ the occurrence of $\cdot$ with $A$

We say that $E$ {\bf closes} $A$ if $E[A]$ is a EP.
For every EP of the form $\{\subst{e}{x}\}|A$, the notation $A^{\{ \subst{e}{ x}\}}$
denotes the EP obtained from $A$ by replacing every free variable
$x$ in $A$ on the term $e$.
It is easy to prove that $A^{\{\subst{e}{x}\}}$ is a correct EP.

\subsection{Equational theories}

We assume that we are given an
{\bf equational theory},
that is, a congruence $\sim$ on terms that is closed under
substitutions of terms instead of variables.
We write $\vdash e = e'$ when
$e\sim e'$.
The notation $\vdash e \neq e'$
means the negation of the statement
$\vdash
e = e'$.
If $\vdash e = e'$, then we will consider the terms $e$ and $e'$ to be the same.

\section{Structural equivalence of processes}

A {\bf structural equivalence} is a smallest equivalence 
$\equiv$ on  EPs such that
\bn\i
$A\equiv B\Rightarrow
A|C\equiv B|C$
and $\nu u.A\equiv \nu u.B$,
\i \bn\i \label{dfgvaergfvsdfg}
$A \equiv A \parop \nil$,
$A \parop (B \parop C) \equiv (A \parop B) \parop C$,
$A \parop B \equiv B \parop A$,
 $\Repl P \equiv P \parop \Repl P$,
\i \label{dfgdfgadgs}$\nu n.A\equiv A$, if $n\not\in A$,
$\nu u.\nu v.A \equiv \nu v.\nu u.A$,
\i\label{fgzerfgdfgdfgfdz} $(\nu u.A)|B\equiv \nu u.(A \parop B)$,
\i\label{zdfbhzdeareg}
$\nu x. (\smx |A)\equiv A^{\smx}$,
\i\label{zfdgzreggggge4} $\smx \parop A \equiv \smx \parop A^{\smx}$,
 $ \{\sx e\} \equiv \{ \sx {e'} \}$ if $\vdash e = e'$.
\en\en

Recall that, according to our convention in section \ref{fdgfsdhdr5}, we assume that in each process under consideration, all bound variables or names are renamed so that they do not appear in any other process under consideration. Therefore, in particular,
in the relation \ref{fgzerfgdfgdfgfdz} in the list above,
the symbol $u$ on the left and right sides of this relation
denote different variables or names.\\

\refstepcounter{theorem}
{\bf Theorem \arabic{theorem}\label{nt1}}.
Each CEP $A$ is structurally equivalent to a EP of the form
\be{erfewfasdf}\by
\nu\vect{n}. (\{\subst{\vect{e}}{\vect{x}}\} \parop P),\\\mbox{ where
$\vect{x}=dom(A)$,
$P$ is a EP,
$\fv(P) = \emptyset$, $var(\vect{e}) = \emptyset$, $
\vect{n} \subseteq
\vect{e}$.}
\ey\ee

{\bf Proof.}
Using the rule \ref{fgzerfgdfgdfgfdz} from the definition of
$\equiv$, we can
replace $A$ with a EP of the form $\nu \tilde n.\nu \tilde x.A'$,
where $A'$ is a EP that does not contain binding operations, i.e.,
$A'$ consists of EPs and substitutions jointed by operation $|$.
Let $x\in \tilde x$, i.e., $A$ has the form $\nu \tilde n.\nu {\tilde x'}.\nu x.A'$.
By the definition of a correct EP,
$A'$ contains a substitution of the form $\{\subst{e}{x}\}$,
i.e., $A'\equiv \{\subst{e}{x}\}|A''$.
According to the rule \ref{zdfbhzdeareg},
$A\equiv \nu \tilde n.\nu {\tilde x'}.\nu x. (\{\subst{e}{x}\}|A'')\equiv \nu \tilde n.
\nu {\tilde x'}.(A'')^{\{\subst{e}{x}\}}$.
Thus, using
the transformations associated with the definition of $\equiv$,
$A$ can be transformed to the form
$\nu \tilde n.(\{\subst{e_1}{x_1}\}|\ldots|\{
\subst{e_l}{x_l}\}|Q)$, where
$Q$ is the PP, and
the condition
\re{sadfvaew} is satisfied. Given this condition, the property
\ref{zfdgzreggggge4} from the definition of $\equiv$, and the closedness property of $A$,
it is easy to see that
$A\equiv \re{erfewfasdf}$. Achieve
properties $\tilde n\subseteq \tilde e$
you can do the following:
if
$\exists\,n'\in \vect{n}:n'\not\in \tilde e$,
then $\nu \vect{n}. (\{\subst{\vect{e}}{\vect{x}}\} \parop P)\equiv
\nu (\vect{n}\setminus \{n'\}). (\{\subst{\vect{e}}{\vect{x}}\} \parop (\nu n'.P)).\;\;
\blackbox$\\

If $A$ is a CEP of the form \re{erfewfasdf},
then
the components $\vect n$ and
$\{\subst{\vect{e}}{\vect{x}}\}$ in this CEP
will be denoted by
$\vect n_A$ and $\theta_A$, respectively.

\section{Actions and transitions}

\subsection{Actions}

Denote by $Act$ the set of {\bf actions},
each of which has one of the following types
(below $c,e,e'\in Tm$):
\bi
\i $c(e)$ and $\bar c\l{e}$
(input and output, respectively, of
message $e$ through channel $c$),
\i $[\![e=e']\!]$ and $[\![e\neq e']\!]$ (testing the condition $e=e'$ or $e\neq e'$).
\ei
Actions of the form $c(e)$ and $\bar c\l{e}$
are called {\bf external actions},
while actions $[\![e=e']\!]$ and $[\![e\neq e']\!]$
are called {\bf internal actions}. The set of external
actions is denoted by $Act^\bullet$.

If an action $\alpha$
has the form $c(e)$ or $\bar c\l{e}$,
then $\bar\alpha$ denotes
the action $\bar c\l{e}$ or $c(e)$
respectively.

\subsection{Transitions}

Each action $\alpha\in Act$
defines a binary relation on the EP,
called a
{\bf transition relation} associated with $\alpha$. If a pair $(A,A')$ belongs to 
this relation, then we will
denote this fact by the notation
$A \ltr{\alpha} A'$, which is called a
{\bf transition}, and which
can
be interpreted as a statement that $A$
can perform the action $\alpha$ and then behave like $A'$.

The rules defining transitions are presented below.
Some transitions are defined
explicitly, while others
defined in the form of inference rules.

Each inference rule
states that
if the statements above the line are true, then the statement below the line also is true,
provided that the EPs included in it are correct.

Below 
 $c,e,e'\in Tm$, $x\in Var$,
$u\in{\cal U}$,
$P,P'$ are PPs,
$A,A', B,B'$ are EPs,
$\alpha\in Act$.

Explicit transitions:
\bi\i $
\Rcv{c}{x}.P
\ltr{ \Rcv{c}{e} }
P^{\{\subst{e}{x}\}}$,
\i $\Snd{c}{e}.P \ltr{ \Snd{c}{e} } \{\subst{e}{x}\} \parop P$,
where $x$ is a {\bf new variable}, i.e. a
variable that has not been occurred
in other EPs considered to the present moment,
\i if $A=\IfThenElse{e}{e'}{P}{P'}$,
then
$A \ltr{[\![e=e']\!]}P$ and
$A \ltr{[\![e\neq e']\!]}P'$.
\ei

Inference rules:
\be{dfgbxhgxnjhfyjggxg}\by
\infrule{
A \ltr{\alpha} A' }{
A \parop B \ltr{\alpha} A'\parop B },
\hspace{3ex} 
\infrule{
A \ltr{\bar c\l{e}} A' \hspace{3ex}
B\ltr{c'(e')}B'
}{
A \parop B \ltr{[\![(c,e)=(c',e')]\!]} A'\parop B'},
\hspace{3ex} 
\infrule{
A \ltr{c(e)} A' \hspace{3ex}
u \not\in c(e)
}{
\nu u.A \ltr{c(e)} \nu u.A'},
\\\;\\
\infrule{
A \ltr{\bar c\l{e}} A' \hspace{3ex}
u \not\in c
}{
\nu u.A \ltr{\bar c\l{e}} \nu u.A'},
\hspace{3ex} 
\infrule{
A \equiv B \hspace{3ex} B \ltr{\alpha} B' \hspace{3ex} B' \equiv A' }{
A \ltr{\alpha} A'}.\ey\ee

\subsection{Transitions
on closed extended processes}

We will assume that
if $A$ is a EP,
then only those
transitions
$A\ltr{\alpha}A'$ will be considered in which
$var(\alpha)\subseteq dom(A)$,
where $var(\alpha)$ is the set of variables
occurred in $\alpha$.

For any CEP $A,A'$ \bi\i
$A\to A'$ means that
either
$A\ltr{[\![e=e']\!]}A'$ and
$\vdash e^{\theta_A}=(e')^{\theta_A}$,\\
or $A\ltr{[\![e\neq e']\!]}A'$ 
and
$\not\vdash e^{\theta_A}=(e')^{\theta_A}$.
\i $A\tot A'$ means that
$\exists\,A_1,\ldots, A_k$:
$A=A_1$, $A'=A_k$, and
$A_i\to A_{i+1}$ if $1\leq i\leq k-1$,
\i $\forall\,\alpha\in Act^\bullet$
$A \ltr{*\alpha*} A'$ means that
$\exists\,B,B':
A\tot B, B\ltr{\alpha} B', B'\tot A'.$
\ei

\section{Observational and labeled equivalences
}\label{sec:ope-sem}
In 
this section we introduce observational and labeled equivalences
on the CEPs. These equivalences allow us to express a wide range of different properties of EPs as statements
about the equivalence of certain CEPs.

We denote by ${\cal M}$ the set of all
binary relations $\mu$ on the set ${\cal P}$ of all CEPs,
with the following properties:
\bi\i if $(A,B)\in \mu$,
then $dom(A)=dom(B)$, and
\i if $(A,B)\in \mu$,
and $A\equiv A'$, $B\equiv B'$, then $(A',B')\in \mu$.\ei

Below, we will assume that all binary 
relations under consideration on ${\cal P}$ belong to ${\cal M}$.

\subsection{Observational equivalence}
\label{sec:equivalences}

If $A$ is a EP, and $a\in Names$, then the notation
$A\barb{a}$ means that
$$\exists\,A', A'',
\exists\,e\in Tm: A\ltr{*} A'\ltr{\bar {a}\l{e}} A''.$$

{\bf Observational bisimulation (OBS)} is a symmetric
relationship $\rel$ on EPs such that if $(A,B)\in {\mu}$, then
\begin{enumerate}
\item $\forall\,a\in Names\;\;A \barb{a}\;\Leftrightarrow\;B \barb{a}$;
\item if $A \ltr{} A'$, then
$\exists\,B'$:
$B \ltr{*} B'$ and $(A', B')\in {\mu}$;
\item for every context $E$ such that
$E[A]$ and $E[B]$ are CEPs, the property
$(E[A], E[B])\in {\mu}$ holds.
\end{enumerate}

\refstepcounter{theorem}
{\bf Theorem \arabic{theorem}\label{naibbimod}}

There is a largest relation on CEPs that
has the properties listed in the definition of
OBS.\\

{\bf Proof}.

The proof below is a slight modification of the proof of a similar statement presented in \cite{htcsmilner}.

Define a function $':{\cal M}\to {\cal M}$,
which maps each  $\mu\in {\cal M}$
to a relation $\mu'\in{\cal M}$, defined as follows:
$$\mu'
\eam \{ (A,B)\in {\cal P}\times {\cal P}\mid
\mbox{ the conditions from the definition of OBS hold}
\}.
$$

Obviously, the function $'$ is monotone, i.e.,
if $\mu_1\subseteq \mu_2$,
then $\mu'_1\subseteq \mu'_2$.

It is easy to see that the relation
$\mu\in {\cal M}$ is a OBS if and only if it is symmetric and
$\mu\subseteq \mu'$.

Consider the set of relations \be{lfgsdougsdfghweui}
\{\mu\in {\cal M}\mid \mbox{$\mu$ is symmetric and }
\mu\subseteq \mu'\}.\ee

Note that the set
\re{lfgsdougsdfghweui} is nonempty, since it
contains, for example, the relation
$\{(A,A)\mid A\in {\cal P}\}$.
Define $\mu_{max}\eam
\bigcup_{\mu\in \re{lfgsdougsdfghweui}}\mu$.

Prove that
$\mu_{max}\in $
\re{lfgsdougsdfghweui}.
$\forall\,\mu\in
\re{lfgsdougsdfghweui}$
from
the inclusion $\by\mu\subseteq \bigcup_{\mu\in\re{lfgsdougsdfghweui}}\mu = \mu_{max}\ey$ and
monotonicity of the function $'$, it follows that $\forall\,\mu\in\re{lfgsdougsdfghweui}\,\mu\subseteq \mu'\subseteq
\mu'_{max}$,
therefore
$\mu_{max}=\bigcup_{\mu\in\re{lfgsdougsdfghweui}}\mu\subseteq \mu'_{max}$,
i.e. $\mu_{max} \in \re{lfgsdougsdfghweui}$.

Thus, $\mu_{max}$ is the greatest element of the set \re{lfgsdougsdfghweui}.
$\blackbox$\\

\refstepcounter{theorem}
{\bf Theorem \arabic{theorem}\label{tretjeuslo32vi223e}}.
$\mu_{max}$ is the greatest fixed point of the function $'$. \\
{\bf Proof}.
The inclusion $\mu_{max} \subseteq \mu'_{max}$
and monotonicity of the function $'$ imply that
$\mu'_{max} \subseteq \mu''_{max}$,
i.e. $\mu'_{max}\in \re{lfgsdougsdfghweui}$,
which, since $\mu_{max}$ is maximal,
implies 
$\mu'_{max} \subseteq \mu_{max}$.
Thus,
$\mu_{max} = \mu'_{max}$.
$\blackbox$\\

\refstepcounter{theorem}
{\bf Theorem \arabic{theorem}\label{tretjeuslo32vie}}.
$\mu_{max}$ is an equivalence.

{\bf Proof}.
\bi
\i $\mu_{max}$ is reflexive, since
$\{(A,A)\mid A\in {\cal P}\} \in \re{lfgsdougsdfghweui}$,
\i $\mu_{max}$ is symmetric, since
it is a union of symmetric
relations,
\i $\mu_{max}$ is transitive, since
If $\mu_1,\mu_2\in \re{lfgsdougsdfghweui}$,
then $\mu_1\circ \mu_2\in \re{lfgsdougsdfghweui}$, therefore
$
\mu_{max}\circ\mu_{max}\in
\re{lfgsdougsdfghweui}$, which implies
$\mu_{max}\circ\mu_{max}
\subseteq \mu_{max}.
\;\;\blackbox
$
\ei

{\bf Observational equivalence} is the relation $\mu_{max}$ defined above.
This relation is denoted by $\bicong$.\\

\refstepcounter{theorem}
{\bf Theorem \arabic{theorem}\label{tretjeuslovie}}.
Let $\mu$ be an equivalence relation on the set of all CEPs.
Then $\mu$ satisfies
the third condition in the definition of OBS
if and only if
\be{zfdgreagfd}\by
\mbox{if
$(A,B) \in \mu$, then
for each CEP $C$ and each list
$\tilde u$ of variables or names, }\\
\mbox{such that $\nu \tilde u.(A|C)$ is CEP,}\quad
(\nu \tilde u.(A|C),\nu \tilde u.(B|C))\in{\mu}.
\ey\ee

{\bf Proof}.

Obviously, \re{zfdgreagfd} follows from the third condition in the definition of 
 OBS.

Prove the converse. Suppose that \re{zfdgreagfd} holds,
$(A,B)\in \mu$,
and $E$ is a context such that $E[A]$ and $E[B]$ are CEPs.
By induction on the number of parallel composition operations in $E$,  prove that \be{fgadgzfdsfdzsf}(E[A],E[B])\in \mu.\ee

If $E$ does not contain a process variable $\cdot$,
then $E[A]=E[B]$, and \re{fgadgzfdsfdzsf} follows from the reflexivity of $\mu$.
If $E=\cdot$, then $(E[A],E[B])=(A,B)\in \mu$.

Consider the remaining case: $E$ contains
an occurrence of $\cdot$ and parallel composition operations.
Using the rule \ref{dfgdfgadgs} from the definition of
$\equiv$, we can move all binding operations outward, i.e.,
replace $E$ with a structurally equivalent context of the form $\nu \tilde u
.
(\cdot|C)$ (which we will also denote by $E$),
where $C$ is a EP.
Thus, $E[A]=\nu \tilde u.(A|C)$.
In this case \re{fgadgzfdsfdzsf}
 follows from \re{zfdgreagfd}.
$\blackbox$

\subsection{Labeled equivalence}

{\bf Labeled bisimulation (LBS)} is a symmetric
$\mu\;\in {\cal M}$, such that
if $(A, B)\in {\mu}$, then
\begin{enumerate}

\item if $A \rightarrow A'$, then
$\exists\,B'$:
$B
\ltr{*} B'$ and $(A', B')\in {\mu}$,
\item $\forall\,\alpha\in Act^\bullet$
if $A \ltr{\alpha} A'$,
then $\exists\,B'$:
$B
\ltr{*\alpha*} B'$ and
$(A', B')\in {\mu}$.
\end{enumerate}

It is easy to prove that theorems \ref{naibbimod},
\ref{tretjeuslo32vi223e}, and
\ref{tretjeuslo32vie} hold for  LBS, i.e.
there is a greatest
relationship which has  properties
listed in the definition of  LBS,
and this relation is an equivalence.
This relation is called a
{\bf labeled equivalence}
and is denoted by
$\wkbisim$.

\section{Coincidence of $\approx$ and $\approx_l$}

In this section, we will prove that
$\approx\,=\,\approx_l$.
This follows from 
$\approx\;\subseteq \;\approx_l$ and $\approx_l\;\subseteq \;\approx$.

\subsection{Proof of the inclusion
$\approx\;\subseteq \;\approx_l$}

To prove 
$\approx\;\subseteq \;\approx_l$
we  prove that $\approx$ is  LBS,
i.e.
if $A\approx B$, then the following statements are true:
\begin{enumerate}



\item If $A \rightarrow A'$, then
$\exists\,B'$:
$B
\ltr{*} B'$ and $A'\approx B'$.
This property follows from the assumption $A\approx B$.
\item If $A \ltr{\alpha} A'$, where
$\alpha\in Act^\bullet$
then $\exists\,B'$,
$B
\ltr{*\alpha*} B'$
and $A'\approx B'$.
\bn\i
Let $\alpha=c(e)$.
Define
$\left\{\by
A_1=A|\bar c\l{e}. a(x){\bf 0}.|\bar a\l{0}.{\bf 0},\\
B_1=B|\bar c\l{e}. a(x).{\bf 0}|\bar a\l{0}.{\bf 0},\ey\right.
\mbox{ where $a$ is a name, $a\not\in A,B$. }$

Since $A\approx B$ implies
$A_1\approx B_1$,
and the definition of $A_1$ implies the property
$A_1\ltr{*} A'|{\bf 0}|{\bf 0}\equiv A'$,
then $\exists\,B_1': B_1\ltr{*} B_1'$, $A'\approx B_1'$.

The property $B_1\ltr{*} B_1'$ is possible for two reasons:
\bi\i $\exists\,B': B\ltr{*} B'$ and $B_1\ltr{*}
B'|\bar c\l{e}. a(x).{\bf 0}|\bar a\l{0}.{\bf 0}=
B_1'$, in this case
$B'_1\Downarrow a$,
$A'\not\Downarrow a$,
which contradicts the assumption $B_1'\approx A'$,
i.e., this case is impossible,
\i $\exists\,B': B\ltr{*c'(e')*}B'$
and $(c,e)^{\theta_B}=(c',e')^{\theta_B}$,
i.e. $B\ltr{*c(e)*}B'$,
in this case
$B_1\ltr{*} B'_1 =
B'|{\bf 0}|{\bf 0}
\equiv B'\approx A'$.
\ei

Thus, in the case  $A \ltr{c(e)} A'$, statement 3 holds.

\i Let $\alpha=\bar c\l{e}$.
This case is considered similarly to the previous one, for which
EPs are 
$$
\left\{\by
A_1=A|c(x). \IfThen{x}{e}{a(y).{\bf 0}}
|\bar a\l{0}.{\bf 0},\\
B_1=B|c(x). \IfThen{x}{e}{a(y).{\bf 0}}|\bar a\l{0}.{\bf 0},
\ey\right.
\mbox{ where $a$ is a name, $a\not\in A,B$.
$\blackbox$}$$

\en
\end{enumerate}

\subsection{Proof of inclusion
$\approx_l\;\subseteq \;\approx$}

To prove the inclusion
$\approx_l\;\subseteq\; \approx$
we  prove that $A\approx_lB$ is OBS,
i.e.
if $A\approx_l B$, then each of the three statements listed below holds.
\begin{enumerate}
\item $\forall\,a\in Names\;\;A \barb{a}\;\Leftrightarrow\;B \barb{a}$.
This  follows from properties 2 and 3 of the definition of  OBS.
\item If $A \to A'$, then
$\exists\,B'$:
$B \ltr{*} B'$ and $A'\approx_l B'$.

This property follows from the assumption $A\approx_l B$
and property 2 from the definition of  LBS.
\item
For every EP $C$ and every list
$\tilde u$ of variables or names,
such that $A_1\eam\nu \tilde u.(A|C)$ is a EP, the property
$A_1\approx_l B_1$ holds, where
$B_1\eam\nu \tilde u.(B|C)$.

To prove this statement, we will prove that the relation
$$\mu\eam \{(\nu \tilde u.(A|C),\nu \tilde u.(B|C))
\in {\cal P}\times {\cal P}
\mid A\approx_l B,
\mbox{ $C$ -- EP}\}$$
is a LBS, i.e.
$\forall\,(A_1,B_1)=
(\nu \tilde u.(A|C),\nu \tilde u.(B|C))\in \mu
$ the following properties hold:
\begin{enumerate}
\item \label{afdasdf2}
if ${A_1} \rightarrow A_1'$, then
$\exists\,B_1'$:
$B_1
\ltr{*} B_1'$, $(A_1',B_1')\in \mu$,
\item \label{afdasdf3}
if $A_1\ltr{\alpha} A_1'$,
where $\alpha\in Act^\bullet$,
then $\exists\,B_1'$:
$B_1
\ltr{*\alpha*} B_1'$, $(A_1', B_1')\in \mu$.
\end{enumerate}

\end{enumerate}


Proof of  
\ref{afdasdf2}: the property
${\nu \tilde u.(A|C)} \rightarrow A_1'$
is possible in one of the following
three cases:
\bi
\i $\exists\,A':A\to A'$, $A_1'={\nu \tilde u.(A'|C)}$,
in this case, from $A\approx_l B$ it follows that
$\exists\,B':B\tot B'$,
$A'\approx_l B'$, therefore
we can define $B_1'\eam
{\nu \tilde u.(B'|C)}$,
\i $\exists\,C':C\to C'$, $A_1'={\nu \tilde u.(A|C')}$,
in this case, $B_1'\eam
{\nu \tilde u.(B|C')}$,
\i $\exists\,A':A\ltr{\alpha} A'$,
$\exists\,C':C\ltr{\bar \alpha} C'$,
$A_1'={\nu \tilde u.(A'|C')}$,
in this case, from $A\approx_l B$ it follows that
$\exists\,B':B\ltr{*\alpha*} B'$,
$A'\approx_l B'$,
therefore
we can define $B_1'\eam
{\nu \tilde u.(B'|C')}$.
\ei

Proof of  
\ref{afdasdf3}: the property
$\nu\tilde u.(A|C)\ltr{\alpha} A_1'$
is possible in one of the following
two cases:
\bi
\i $\exists\,A':A\ltr{\alpha} A'$, $A_1'={\nu \tilde u.(A'|C)}$,
in this case, from $A\approx_l B$ it follows that
$\exists\,B':B\ltr{*\alpha*} B'$,
$A'\approx_l B'$, therefore
we define $B_1'\eam
{\nu \tilde u.(B'|C)}$,
\i $\exists\,C':C\ltr{\alpha} C'$, $A_1'={\nu \tilde u.(A|C')}$,
in this case, $B_1'\eam
{\nu \tilde u.(B|C')}$. $\blackbox$
\ei

\section{Conclusion}

The main result of this paper is a
simplified proof
of one of the fundamental results of applied pi-calculus:
the theorem that the concepts of
observational and labeled equivalence
of extended processes in applied pi-calculus coincide. 

One of the problems for further
research in this area
is to find algorithms for checking observational equivalence for sufficiently broad classes of pi-calculus processes. One approach to solving this problem is to
represent the analyzed processes as graphs whose edge labels are actions from the set $Act$,
and the proof of observational equivalence can consist
of graph reduction for the analyzed processes and proving
isomorphism of the reduced graphs.

\end{document}

\begin{abstract}
В настоящей работе излагается новая
математическая модель криптографических
 протоколов, и приводятся примеры применения
 этой модели для решения задач верификации
 криптографических протоколов. 
Криптографические протоколы -- это коммуникационные протоколы, 
реализованные с применением криптографических  алгоритмов 
для решения задач защиты  информации, в рамках которого стороны  
информационного взаимодействия  последовательно выполняют 
определенные действия и обмениваются сообщениями.
Они используются, 
например, в 
электронных платежах, электронных 
процедурах голосования, 
системах доступа 
к конфиденциальным данным, 
и т.д. Ошибки в криптографических
 протоколах
могут привести к 
большому ущербу, 
поэтому необходимо 
использовать 
математические методы для
обоснования различных свойств корректности
и безопасности криптографических
 протоколов. 
В работе излагаются новые методы формальной верификации 
криптографических протоколов. 
Для моделирования криптографических протоколов
 в работе вводятся понятия последовательного и распределенного процессов.
Особенностью модели протоколов является её простота по сравнению 
с другими моделями протоколов, основанных на логических формулах 
или на алгебраических процессных выражениях. Участники протоколов
представляются в виде графов, представляющих системы переходов.
Действия,  выполняемые участниками, являются метками этих переходов. 
Методы обоснования корректности протоколов, рассматриваемые в настоящей статье, связаны с рассуждениями
для графов, которые более просты и наглядны по сравнению с методами,
основанными на построении логического вывода в логических и алгебраических моделях протоколов.\\

{\bf Ключевые слова}: криптографические протоколы, 
последовательные процессы, 
распределенные процессы, верификация
\end{abstract}

\section{Введение}

\subsection{Понятие криптографического протокола}

{\bf Криптографический протокол (КП)}  --
это коммуникационный протокол, 
реализованный с применением криптографических  алгоритмов 
для решения задач защиты  информации, в рамках которого стороны  
информационного взаимодействия  последовательно выполняют 
определенные действия и обмениваются сообщениями.
КП  представляет собой  распределенный алгоритм, 
 описывающий порядок обмена 
сообщениями  между несколькими агентами.
Примеры таких агентов -- компьютерные
системы, банковские карточки, люди, и т.д.

Для обеспечения свойств безопасности КП
(таких например как конфиденциальность
передаваемых данных)
в КП могут использоваться криптографические преобразования
(шифрование, электронная подпись, 
хэш-функции, и т.п.). 
Мы предполагаем, что 
криптографические преобразования, 
используемые в КП, являются 
идеальными, т.е. удовлетворяют
некоторым аксиомам, выражающим, 
например, невозможность 
извлечения открытых текстов из 
шифртекстов без знания 
соответствующих
криптографических ключей.

\subsection{Уязвимости в криптографических
протоколах}

Многие уязвимости в КП связаны
 не с плохими криптографическими качествами 
используемых в них криптографических примитивов, 
а с логическими ошибками в КП. 
Наиболее ярким примером уязвимости в КП является
уязвимость в КП аутентификации Нидхэма-Шредера \cite{needhams}, который был опубликован в 1978 г., и использовался  в критических по безопасности информационных системах. Cпустя более  16 лет после начала  использования этого КП
в нем обнаружилась логическая ошибка \cite{r2332}, 
связанная с возможностью непредусмотренного 
нечестного
поведения одного из участников этого КП
и нарушающая свойство безопасности этого КП.
Особенность этой ошибки заключается в том, что 
данный КП является предельно простым распределенным алгоритмом, состоящим всего из трех действий, и при визуальном анализе этого КП 
отсутствие в нем ошибок
не вызывало никаких
сомнений. Ошибка была обнаружена лишь
при помощи инструмента автоматизированной 
верификации КП. 

Другой пример логической ошибки в КП (взят из 
статьи \cite{mypapersverc}):
в КП  входа в портал Google, 
позволяющем пользователю
идентифицировать себя только один раз, а затем обращаться к различным приложениям
(таким, например, 
как Gmail или календарь Google), 
обнаружена логическая ошибка, позволяющая 
нечестному поставщику услуг выдавать себя за любого из своих пользователей для другого
поставщика услуг.

Существует много других 
примеров КП
(см. например 
\cite{mypapers1}-\cite{mypaperskerberos}),
в которых обнаружились
уязвимости следующего вида:
\bi
\i  участники этих КП
могут получать  искаженные
сообщения (или вообще терять их)
в результате перехвата, 
удаления или искажения
противником
передаваемых сообщений, 
что нарушает
свойство целостности передаваемых сообщений,
\i противник может узнать 
секретную информацию,
содержащуюся в перехваченных
сообщениях, 
в результате чего нарушается 
свойство конфиденциальности
передаваемых сообщений.
\ei

Также есть примеры уязвимостей
в КП, используемых  для аутентификации перед провайдерами мобильной телефонной связи, для снятия  денег в банкомате, для работы с электронными паспортами, проведения электронных выборов, и т.д.

Все эти
примеры являются обоснованием
того, что в критических по безопасности системах недостаточно неформального анализа требуемых свойств безопасности используемых в них
КП,
необходимо 
\bi
\i построение
{\bf математических моделей}
  анализируемых КП,
\i описание  свойств
анализируемых КП в виде математических объектов, называемых
{\bf  спецификациями} свойств этих КП, и 
\i построение формальных
  доказательств утверждений о том, что 
анализируемые КП 
удовлетворяют (или не удовлетворяют)
своим спецификациям, процедура построения таких 
доказательств называется
{\bf  верификацией} анализируемых КП.
\ei

В настоящей работе строится новая 
математическая модель КП, в терминах которой
можно выражать такие свойства корректности 
КП, как например 
конфиденциальность
 передаваемых сообщений
(т.е. обоснование следующего свойства анализируемого КП:
содержание сообщений,
посланных одним участником этого КП другому
участнику этого КП, 
не будет известно противнику), 
или 
аутентификация (т.е. доказательство
подлинности) участников КП.

\subsection{Основные методы
моделирования и 
верификации криптографических
протоколов}

Обзоры наиболее широко используемых 
методов моделирования и верификации КП
содержатся в книгах
\cite{C324M6} и \cite{mypa23persstrandlast}. 
Основные классы 
моделей КП и подходов к верификации
КП имеют следующий вид.
\bn
\i {\bf Логические модели}.

Данный класс моделей был самым первым
подходом к моделированию и верификации
КП. На основе данного класса моделей
проблема
 верификации КП сводится к проблеме
 построения в некотором логическом исчислении
доказательства 
 теоремы о том, что анализируемый КП
обладает заданными свойствами. 
В работе \cite{mypapers5}
была изложена первая математическая модель КП, 
называемая {\bf логикой BAN}
(название этой логики соответствует
фамилиям ее создателей -- Бэрроуза,
Абади и Нидхэма).
Данная модель имеет большие ограничения:
в ней предполагается, что участники 
анализируемого КП являются честными,
т.е. точно выполняют предписания КП. 
Такое ограничение не позволило обнаружить
упомянутую выше уязвимость в КП Нидхэма-Шредера.
Кроме того, данная модель
не позволяет анализировать КП 
с неограниченным порождением
сеансов. Аппарат логики BAN был развит в работах
\cite{CM87}-\cite{CM151}.
Важным классом логических исчислений 
для моделирования и анализа КП является
композиционная логика протоколов 
(Protocol Composition Logic), которой посвящены
работы 
\cite{CM78}-\cite{CM67}.
Одним из классов логических моделей КП 
связан с логическим программированием. 
В данных моделях 
шаги протокола представляются в виде правил 
переписывания термов.
Для моделирования КП используются 
клаузы Хорна и системы 
уравнений с ограничениями (constraint systems).
Данный подход
излагается в работах 
\cite{mypapershartog4}, \cite{mypapershartog3}  и др.

Важным классом
логических методов моделирования
и анализа КП является
индуктивный метод Паульсона:
\cite{CM130}-\cite {8}.

\i Модели, основанные на {\bf алгебре процессов}.

Источником данного класса моделей является
основополагающая работа Р.Милнера 
\cite{milner}. В данной работе строится
модель взаимодействующих 
процессов, в которой процессы 
представляются термами. На этих термах
вводится отношение наблюдаемой эквивалентности,
которое позволяет эффективно выражать 
различные свойства 
процессов, связанные с безопасностью
(в частности свойства секретности и анонимности).
Первой работой, в которой излагается 
модель КП на базе подхода Р.Милнера, 
является статья М.Абади и А.Гордона 
\cite{CM3}. Среди других работ, относящихся 
к этому направлению, можно отметить работы
\cite {82}-\cite{n1}.

\i Модели, основанные на  
{\bf CSP}.

{\bf CSP} ({\bf  Communicating Sequential Processes}) --
это математический аппарат, разработанный А.Хоаром 
\cite{41} и предназначенный для моделирования
и анализа
распределенных вычислительных процессов.
На базе этого аппарата построен метод моделирования
и верификации КП, наиболее полно изложенный в книге 
\cite{CM1321}. Дедуктивная
верификация КП на основе данного подхода 
использует понятие {\bf ранг-функции}.
Среди работ, относящихся к данному направлению,
можно отметить работы
\cite{88}-\cite{n324}.

\i Модели, основанные на {\bf пространствах нитей}
({\bf strand spaces}).

Пространства нитей позволяют представлять
процессы, входящие в КП, 
в виде графических объектов (называемых
нитями), в которых
указаны зависимости между действиями,
относящимися к различным процессам. 
Среди работ, относящихся к методам
моделирования и верификации
КП на основе понятия пространства нитей, 
можно отметить работы
\cite{CM153}-\cite{mypapersae2bf}.
\en

\subsection{Сравнение предлагаемой модели
криптографических протоколов
с другими моделями}

Модель КП, излагаемая в настоящей работе,
унаследовала наиболее существенные качества моделей
каждого из перечисленных выше четырех классов.
В этой модели КП 
представляются
в виде 
распределенных процессов (РП), 
взаимодействующих
путем асинхронной
передачи сообщений через каналы.
Каждый РП, соответствующий
какому-либо КП, 
 представляет собой совокупность
последовательных процессов (ПП),
моделирующих работу участников этого КП.
Как правило, \bi
\i эти ПП представляют собой 
последовательности действий, 
которые графически можно изобразить
в виде нитей, и \i выполнение всего КП
можно представить в виде пространства нитей,
точки на которых связаны ребрами, 
изображающими передачу и прием
сообщений.\ei

Свойства КП могут представляться
в виде логических формул, для обоснования которых
могут использоваться стандартные алгоритмы логического вывода. 
Кроме того, некоторые свойства КП (например анонимность) м.б. выражены в виде отношения
наблюдаемой эквивалентности между 
соответствующими РП,
аналогично тому, как это делается в моделях КП 
основанных на процессной алгебре.

Основное достоинство предложенной модели заключаются в том, что
доказательства свойств корректности КП  на основе данной модели
м.б. существенно короче, 
чем доказательства этих свойств
на основе других моделей КП. 
Для обоснования этого  
утверждения мы приводим
пример верификации 
КП Yahalom  \cite{CM1321}.
Верификация этого КП
в вышеупомянутом источнике 
занимает несколько 
десятков страниц, в то время как 
верификация КП Yahalom  на базе предложенной
модели 
занимает менее 4 страниц.
Причина такого существенного упрощения   
верификации КП связана с использованием
доказанных в настоящей работе теорем  \ref{t1} и \ref{t2}, которые
представляют собой схемы обоснования свойства защищённости 
сообщений при выполнении переходов протокола,
и устраняют дублирование рассуждений при доказательстве корректности КП.

Предложенный формализм предназначен для построения формальных 
доказательств корректности таких КП, которые являются корректными.
Он не предназначен для обнаружения ошибок в некорректных КП.
Если протокол является некорректным, то для обнаружения ошибок в нем
должен использоваться другой метод, аналогичный методу Model Checking
в теории верификации программ, см. \cite{modelchec}.
Данный метод будет изложен  в последующих публикациях автора.

Изложенный в статье
язык описания РП имеет самостоятельную 
ценность, и может рассматриваться 
как новый язык описания распределенных 
алгоритмов с применением криптографических функций.

В целях простоты изложения в работе рассматривается такая математическая модель КП, которая отражает простейшие
криптографические примитивы,
используемые в  КП: в ней формализуются лишь симметричные системы шифрования, и не рассматриваются
такие примитивы как системы шифрования с открытым ключом, 
хэш-функции, цифровые подписи, 
и т.п.
Все эти примитивы несложно ввести в представленную модель путем соответствующих дополнений.

\section{Вспомогательные понятия}

В этом параграфе мы излагаем понятия, 
необходимые для определения понятий
последовательного и распределённого
процесса.

\subsection{Типы, переменные,
константы и функциональные символы}

Будем предполагать, что заданы следующие множества. 

\bi\i 
Множество $Types$, его элементы называются {\bf типами данных}
(или просто {\bf типами}). 
Каждому типу $\tau$ из $Types$ сопоставлено множество
$D_\tau$ {\bf значений} типа $\tau$.
\i  Множества $Var$ и $Con$, их элементы называются {\bf переменными} и
{\bf константами} соответственно. Каждой переменной $x\in Var$ и константе $c\in Con$ сопоставлен тип $\tau(x)$ и $\tau(c)\in Types$ соответственно.
Каждая переменная $x$ может принимать {\bf значения} в некотором множестве, которое будем обозначать $D_{\tau(x)}$,
т.е. в различные моменты времени переменная $x$ может быть связана с различными элементами множества $D_{\tau(x)}$.
\i Множество $Fun$, его элементы называются {\bf функциональными
символами (ФС)}. Каждому $f\in Fun$ сопоставлен {\bf функциональный тип (ФТ)} $\tau(f)$, который представляет собой запись вида
\be{l1}
(\tau_1, \ldots, \tau_n) \to \tau,\mbox{ где }\tau_1, \ldots, \tau_n, \tau \in Types.\ee
\ei

Будем считать, что среди типов, входящих в указанное выше множество
$Types$, присутствуют следующие типы:
\bi\i {\bf A}, значения этого типа называются {\bf агентами},
\i  {\bf K}, значения этого типа обозначают {\bf ключи}, используемые агентами для шифрования или расшифрования сообщений,
\i {\bf M}, значения этого типа обозначают {\bf сообщения}, которые агенты могут пересылать друг другу во время своей работы,
\i для каждого типа $\tau$ множество
$Types$ содержит тип ${\bf 2}^\tau$, значениями которого являются подмножества 
множества $D_\tau$,
тип ${\bf 2}^\tau$ используется, например, для 
представления содержимого канала: его значениями могут быть 
множества сообщений, содержащихся в канале, 

\i для каждого списка типов $\tau_1, \ldots, \tau_n$ 
множество $Types$ содержит тип $(\tau_1, \ldots, \tau_n)$,
и $D_{(\tau_1,\ldots,\tau_n)} = D_{\tau_1} \times\ldots
\times D_{\tau_n}$.
\ei

\subsection{Термы}\label{termy}

{\bf Термы} строятся из переменных, констант и ФС. Множество всех термов
обозначается символом $Tm$. Каждый терм $e$ имеет тип 
$\tau(e)\in Types$,
определяемый структурой терма $e$.

Правила построения термов имеют следующий вид:

  \bi
  \i если $x\in   {\it Var}\cup Con$,
то $x$ --  терм типа $\tau(x)$, и
\i 
если $e_1,\ldots, e_n\in Tm$,
$f\in Fun$, и
$\tau(f)$ имеет вид \re{l1}, где 
$\tau_i=\tau(e_i)$
$(i=1,\ldots, n)$,
то запись $f(e_1,\ldots, e_n)$ --
терм типа $\tau$.
\ei

Терм $e\in Tm$ называется {\bf подтермом} терма $e'\in Tm$,
если либо $e=e'$, либо $e'=
f(e_1,\ldots, e_n)$, 
и $\exists\,i\in \{1,\ldots, n\}$:
$e$ -- подтерм терма 
$e_i$.
Запись $e\subseteq e'$, где $e,e'\in Tm$,
 означает, что $e$ является
  подтермом терма $e'$.
Запись $e\subset e'$, где $e,e'\in Tm$,
 означает, что $e\subseteq e'$ и $e\neq e'$.
 
 Индукцией по структуре терма $e$
 нетрудно доказать, что
 \be{l2}
\by
\mbox{если $e_1$ и $e_2$ --  различные подтермы
 терма $e$,} \\\mbox{то 
  либо $e_1\subset e_2$,
либо $e_2\subset e_1$,}\\\mbox{либо 
 $e_1$ и $e_2$ не имеют общих компонентов.}
  \ey
 \ee

Будем считать, что $Fun$ содержит следующие ФС:
\bi\i ФС $encrypt$ типа $({\bf K},{\bf M}) \to  {\bf M}$,\\
терм вида $encrypt(k,e)$ обозначает сообщение, получаемое шифрованием сообщения $e$ на ключе $k$
в симметричной системе шифрования, будем обозначать такой
терм записью $k(e)$ и называть его {\bf шифрованным 
сообщением (ШС)},
\i ФС $shared\_key$ типа $({\bf 2}^{\bf A}) \to  {\bf K}$,\\
терм вида $shared\_key(\{A_1, \ldots,A_n\})$ 
называется {\bf разделяемым
ключом} агентов $A_1, \ldots ,A_n$ и будет обозначаться  $k_{A_1, \ldots ,A_n}$,
\i ФС $list$ типа 
$(\tau_1, \ldots, \tau_n) \to  (\tau_1, \ldots, \tau_n)$, 
где $(\tau_1, \ldots, \tau_n)$ -- произвольные типы,
терм вида $list(e_1,\ldots, e_n)$ обозначает список термов, компоненты 
которого -- термы $e_1,\ldots, e_n$,
мы будем обозначать терм  $list(e_1,\ldots, e_n)$
более короткой записью
$(e_1,\ldots, e_n)$. 
\ei

Отметим, что мы не предполагаем наличие в $Fun$
ФС $decrypt$, обозначающего операцию расшифрования
в симметричной системе шифрования.
Это связано с тем, что операцию расшифрования
шифртекста $e$ на ключе $k$
можно выразить 
в нашем языке путем действия вида $k(x):=e$, 
которое 
заключается в нахождении такого значения переменной $x$,
результат
шифрования которого на ключе $k$ будет совпадать с 
значением терма $e$
(понятие действия изложено в пункте \ref{posrasprproc}).

Поясним также,  каким образом в рассматриваемой модели 
осуществляется проверка возможности доступа участника 
к зашифрованной информации.
Согласно определению понятия действия вида $e:=e'$, операция расшифрования может быть выполнена участником в том и только том случае когда ключ, 
который он использует для расшифрования, совпадает с тем ключом, на котором было зашифровано расшифровываемое сообщение. Если эти ключи разные, то расшифрование не может быть выполнено, т.е. выполнение протокола не дойдёт 
до своего заключительного состояния. 

Также отметим, что мы не рассматриваем формализацию в
нашей модели таких компонентов криптографических протоколов 
как асимметричное шифрование,
цифровая подпись, хэш-функции, схемы разделения секрета, 
и т. п. Формализация всех этих компонентов является несложной
и производится
на основе введения дополнительных ФС.

Будем использовать следующие обозначения:
\bi
\i  $\forall\,x\in Var,\forall\, e\in Tm$
запись $x\in e$
означает, что $x$ входит в $e$,
\i  $\forall\,E\subseteq Tm\;\;
Var(E)=\{x\in Var\mid \exists \,e\in E:
 x\in e\},$
\i $\forall\,E\subseteq Tm$
запись $tm(E)$ обозначает 
наименьшее по включению
множество,  удовлетворяющее следующим условиям:
\bi\i
$E\subseteq tm(E)$, 
$Con\subseteq tm(E)$, 
\i для каждого терма вида
$f(e_1,\ldots, e_n)$
\begin{center}
если
$e_1,\ldots, e_n\in tm(E)$, 
то $f(e_1,\ldots, e_n)\in tm(E)$,
\end{center}
\ei
\i $\forall\;\tau\in Types$,
$\forall\,E\subseteq Tm$
$E_\tau=\{e\in E\mid \tau(e)=\tau\}$.

\ei

\subsection{Формулы и теории}

{\bf Элементарной формулой (ЭФ)}
называется запись одного из следующих
видов:
$$\by
e=e',\quad\mbox{где }\tau(e)=\tau(e'),\\
e\in e',\quad\mbox{где }\tau(e') = {\bf 2}^{\tau(e)}.
\ey$$

{\bf Формулой} называется обычная 
булева комбинация
ЭФ, в которой могут использоваться 
   логические связки $\neg$, $\wedge$, $\vee$, $\to$, $\leftrightarrow$
   и константы 1 и 0.
Множество всех формул обозначается  $Fm$.
   Формулы вида $\wedge(e_1,e_2)$
    будем записывать в более привычном виде
   $e_1\wedge e_2$, аналогичная
   запись используется для других
   булевых комбинаций.
   Формулы вида
$e_1\wedge\ldots\wedge e_n$ 
   могут также записываться в виде
   ${\def\arraystretch{0.5}\c{
   e_1\\\ldots\\e_n}}$
или
$\{e_1,\ldots, e_n\}$.
Возможна конъюнкция
произвольного семейства формул
$\{e_i\mid i\in I\}$,
она обозначается записью
 $\bigwedge_{i\in I} e_i$.

{\bf Теорией} называется произвольная
совокупность формул $Th\subseteq Fm$.
Понятие {\bf 
доказуемости} формулы 
в теории определяется стандартным 
образом. Ниже мы будем 
предполагать, что задана некоторая 
теория $Th$, и будем говорить что формула
$\varphi$ {\bf доказуема}, если
она доказуема в этой теории $Th$.

\subsection{Подстановки}
\label{sadrew}

{\bf Подстановкой} называется
 функция 
$\theta:{\it Var}\to Tm$,
такая, что $$\forall\,x\in Var\;\;\tau(x)=\tau(\theta(x)).$$
Будем говорить, что  $\theta$
заменяет переменную 
$x\in {\it Var}$ на терм $\theta(x)$.
Подстановка называется {\bf тождественной}, если 
$\forall\,x\in {\it Var}\;\;\theta(x)=x$.

Будем использовать следующие обозначения:
\bi\i $\Theta$ обозначает 
множество всех подстановок,
\i 
 $\forall\,\theta\in \Theta,\;\forall\,
e\in Tm$ 
запись $e^{\theta}$
обозначает терм, получаемый из $e$
заменой для каждой переменной
$x\in {\it Var}(e)$
каждого вхождения 
  $x$ в $e$ на терм $\theta(x)$;
\i 
 $\forall\,\theta\in \Theta,\;\forall\,
E\subseteq Tm$ 
 $E^{\theta}=\{e^\theta\mid e\in E\}$. 
\i 
$\forall\,\theta\in \Theta,\forall\,\varphi\in Fm$
запись $\theta\vdash \varphi$
обозначает утверждение, что формула
$\varphi^\theta$ доказуема.
\ei




\section{Последовательные 
и распределен\-ные процессы}
\label{posrasprproc}

В этой главе определяются понятия последовательного  и распределенного процесса. 
Последовательный 
процесс является моделью участника КП, а распределенный процесс
является моделью всего КП. Предложенная модель является теоретической основой для метода верификации КП, излагаемого в пункте \ref{l6}.

\subsection{Действия}

{\bf Действие} -- это 
запись 
одного из следующих   видов:
$$\by
\circ!e,\quad \circ?e,\quad e:=e', \quad
\mbox{где } e,e'\in Tm,\ey$$
которые называются
{\bf выводом} сообщения $e$ в 
открытый канал $\circ$, 
 {\bf вводом} сообщения $e$ из 
 открытого канала $\circ$, 
и  {\bf присваиванием},
соответственно. 
Множество всех 
 действий
обозначается  
$Act$. 


\subsection{Последовательные процессы}

{\bf Последовательным процессом} 
(или просто {\bf процессом})
будем называть 
 граф $P$  со следующими
свойствами:
\bi\i  $P$ имеет выделенные
вершины
$\odot$ и $\otimes$, 
называемые {\bf начальной} и {\bf терминальной}
вершинами соответственно, 
из $\otimes$ не выходят рёбра,
\i каждому ребру 
графа $P$ сопоставлена метка
$a\in Act$,
ребро 
процесса $P$ представляется записью
$v\ra{a}v'$, где $v$ и $v'$ --
 начало и конец  ребра,
$a$ -- метка  ребра.

\ei 

Процесс является описанием поведения дискретной
 динамической системы, работа которой
 заключается в последовательном
 выполнении действий, связанных
  с вводом и выводом сообщений
и 
  изменением значений
переменных.
С каждым процессом $P$ связаны 
\bi\i {\bf агент} $Agent_P\in Var_{\bf A}$,
называемый {\bf исполнителем}
процесса $P$,\\
напомним, что согласно обозначениям, введенным в конце
пункта \ref{termy}, 
$Var_{\bf A}$ -- это множество переменных, имеющих тип ${\bf A}$
 (агент),
\i множество 
 $Var_P$ 
{\bf переменных} процесса $P$,
 являющееся
дизъюнктным объединением
следующих множеств:
\bi
\i множество $Public_P$ 
{\bf открытых
переменных}, 
\i множество $Private_P$ 
{\bf приватных
переменных},
\i множество
$Unique_P$ 
переменных, инициализированных уникальными значениями,
эти переменные 
обозначают
криптографические ключи,
или переменные, называемые
{\bf нонсами},
\i $\{x_P\}$, значения $x_P$ --
подмножества множества
$$Public_P\cup Private_P\cup Unique_P,$$
их элементы называются
{\bf инициализированными переменными} процесса $P$,\i
$\{at_P\}$, значения $at_P$ -- вершины
графа $P$,
\i $\{x_\circ\}$, 
 $\tau(x_\circ)={\bf 2}^{\bf M}$, 
 значения $x_\circ$ интерпретируются
 как {\bf содержимое открытого канала}
(отметим, что переменная $x_\circ$ является общей для всех процессов).
\ei
\ei

В каждый момент выполнения процесса каждая 
его неслужебная переменная (т.е.
отличная от $at_P$, $x_P$, $x_\circ$)  либо инициализирована
каким-либо значением, 
либо неинициализирована. Если она 
в какой-либо момент времени инициализирована, 
то во все последующие моменты данная
переменная имеет то же значение что и в момент инициализации.
Переменные из $Unique_P$ инициализированы в начальный момент.

\subsection{Процесс противника}

{\bf Процесс противника} -- это процесс ${\enemy} $, обладающий  свойствами:
\bi 
\i множество вершин графа 
процесса ${\enemy} $ одноэлементно,
\i $\forall\,a\in Act$ граф 
процесса
${{\enemy}}$ содержит ребро
с меткой $a$.
\ei

Ниже будем предполагать, что ${\enemy} $ -- единственный
из всех рассматриваемых 
процессов, граф которого имеет циклы.

\subsection{Распределенные процессы}

{\bf Распределенным процессом (РП)} называется  семейство про\-цессов
${\cal P}=\{P_i\mid i\in I\}$,
таких, что компоненты
семейства 
\be{l3}
\{Private_{P_i}
\cup
Unique_{P_i}\cup
\{x_{P_i}, at_{P_i}\}
\mid i\in I\}\ee
дизъюнктны (если это условие
не выполняется,
то соответствующие переменные 
в процессах $P_i$
переименовываются).

С каждым РП ${\cal P}$
связано 
{\bf начальное состояние}
$\theta_{\cal P}^0\in \Theta$,
обладающее следующими свойствами:
$$\by
\forall\,P\in {\cal P}\quad
x_P^{\theta_{\cal P}^0}= Public_P\cup Unique_P,\;
at_P^{\theta_{\cal P}^0}=\odot, x_\circ^{\theta_{\cal P}^0}=\emptyset.
\ey$$


Если РП 
состоит из одного процесса $P$,
то он обозначается тем же символом $P$.
 Если 
$\{{\cal P}_i\mid i\in I\}$ -- 
семейство РП, то данная запись
обозначает также РП, состоящий из всех процессов, входящих в какой-либо РП из семейства
${\cal P}_i\;(\forall\,i\in I)$.

 Записи  $Public_{\cal P}$ и  $Unique_{\cal P}$
 обозначают   соответственно
 множества
 $$\by\bigcup_{P\in {\cal P}}Public_P \quad \mbox{и}\quad
 \bigcup_{P\in {\cal P}}Unique_P.\ey$$

\subsection{Переходы в распределенных процессах}

{\bf Переход} в РП ${\cal P}$ -- 
это утверждение, обозначаемое записью 
$\theta\ra{a_{P}}\theta'$,
где $P\in {\cal P}$,
 $\theta,\theta'\in \Theta$
($\theta$ называется {\bf началом}
данного перехода, а $\theta'$ -- его {\bf концом}) 
и
 $a$   
-- метка некоторого ребра $v\ra{a}v'$
процесса $P$,
причём выполнены  условия:
\bn
\i $at_{P}^\theta=v$,
$at_{P}^{\theta'}=v'$,
\i
$\forall\,x\in x_{P}^\theta\setminus\{at_{P},
x_{P},
x_\circ\}\;\;x^{\theta}=x^{\theta'}$,

\i если $a=\circ!e$, то 
\bi\i
$e\in tm(x_{P}^\theta),
   x_{P}^{\theta'}=x_{P}^{\theta},
x_\circ^{\theta'}=
x_\circ^{\theta}\cup\{e^\theta\}$,
\i если $e^\theta$ содержит подтерм вида
$k(\tilde e)$, где $k\in Tm_{\bf K}$,
то верно следующее свойство:
\be{l4}
\mbox{либо $k\in Var$, либо
$\left\{\by k=shared\_key(\ldots)\\
Agent_P\in k\ey\right\}$,}
\ee
напомним, что согласно обозначениям, введенным в конце
пункта \ref{termy}, $Tm_{\bf K}$ -- это множество термов, 
имеющих тип ${\bf K}$ (ключ), 
и $Agent_P\in k$ означает, что переменная 
$Agent_P$ входит в терм $k$,
\ei
\i если $a=\circ?e$
или $e:=e'$, то 
\bn
\i $x_\circ^{\theta'}=
x_\circ^{\theta}$,
$x_{P}^{\theta'}=x_{P}^\theta\cup 
Var(e)
$,
\i $\theta'\vdash e\in x_\circ$
или 
 ${\def\arraystretch{0.5}\c{
   \theta'\vdash e=e'
   \\
   e'\in tm(x_{P}^\theta)
   }}$
соответственно,
\i если $e^{\theta}$ содержит подтерм вида 
$k(\tilde e)$, где $k\in Tm_{\bf K}$,
то верно свойство \re{l4},
\i если $a=(e:=e')$ и $k=shared\_key(\ldots)$, то
верна импликация
\be{l5}
k\subseteq (e')^\theta\;\Rightarrow\;
Agent_P\in k.
\ee
\en
\en

Переход $\theta\ra{a_{P}}\theta'$ 
РП ${\cal P}$
 интерпретируется
 как выполнение 
  процессом $P\in {\cal P}$
 действия $a$, в результате чего 
 ${\cal P}$ переходит от 
 $\theta$ к $\theta'$.
 Если в текущий момент с ${\cal P}$
связана подстановка $\theta$,
 и в этот момент
 некоторый процесс $P$, 
 входящий в ${\cal P}$, 
содержит ребро 
 $v\ra{a}v'$,
 причем $v=at_{P}^\theta$, то 
 мы считаем, что РП ${\cal P}$,
связанный в текущий момент
 с подстановкой $\theta$,
 может выполнить действие $a$,
после чего он будет связан 
с подстановкой $\theta'$,
удовлетворяющей
вышеприведённым условиям
при этом
\bi
\i при выполнении действия 
$\circ!e$ происходит добавление
терма $e^\theta$ к содержимому 
открытого канала
$\circ$,
\i при выполнении действия
$\circ?e$ или $e:=e'$
происходит либо чтение некоторого терма из содержимого канала $\circ$,
либо присваивание соответственно,
путем инициализации неинициализированных в текущий момент
переменных из терма $e$:
терм $e$ рассматривается как шаблон,
которому должен соответствовать некоторый терм из $x_\circ^\theta$ или терм $(e')^\theta$
соответственно, и выполняемое действие
заключается в преобразовании $\theta$
в подстановку $\theta'$
путем определения подходящих значений
переменных из $Var(e)\setminus x_P^\theta$,
с таким расчётом, чтобы значение
$e^{\theta'}$ было бы равно некоторому
терму из  $x_\circ^\theta$ или терму $(e')^\theta$ соответственно. 
\ei

\subsection{Выполнение распределенного процесса}

{\bf Выполнение} РП ${\cal P}$ -- это последовательность подстановок
$\pi=(\theta_0,\theta_1,\ldots)$ РП ${\cal P}$, такая, что $\theta_0$ -- начальное состояние РП
${\cal P}$, и 
для каждой пары $\theta_i$, 
$\theta_{i+1}$
соседних членов этой последовательности
имеется переход 
$\theta_i\ra{a_{P}}\theta_{i+1}$,
где $P$ -- 
какой-либо процесс
из ${\cal P}$.

Для каждого выполнения 
$\pi=(\theta_0,\theta_1,\ldots)$
запись $\theta\in \pi$ означает, что
$\exists\,i\geq 0:\theta_i=\theta$.

Если задано выполнение $\pi=
(\theta_0,\theta_1,\ldots)$ и $\theta,\theta'$ -- 
подстановки, входящие в $\pi$, то запись
$\theta<_\pi\theta'$ означает, 
что 
$\theta=\theta_i$ и $\theta'=\theta_j$
для некоторых индексов $i<j$.
Запись
 $\theta\leq_\pi\theta'$ означает, 
что $\theta<_\pi\theta'$ 
или $\theta=\theta'$.

Подстановка $\theta$ РП ${\cal P}$
называется {\bf достижимым состоянием} РП ${\cal P}$, 
если она входит в некоторое выполнение
${\cal P}$.
Множество  всех достижимых состояний РП
${\cal P}$
обозначается  $\Theta_{\cal P}$.

В начальный момент
 выполнения
РП ${\cal P}$
переменные из $Unique_{\cal P}$
инициализированы
{\bf уникальными значениями}, т.е. 
такими значениями, которые
никогда не встречались среди 
всех значений, используемых
до  начала выполнения 
 ${\cal P}$.

\section{Свойство защищённости}

\subsection{Определение свойства защищённости}

В рассуждениях, 
связанных с верификацией
РП, будем
использовать свойство {\bf защищённости}, 
обозначаемое записью 
\be{l6}\mbox{$E\,\bot\, P$,
где $\left\{
\by E\subseteq Public_{\cal P}\cup
tm(Public_{\cal P})_{\bf K},\\
P\in {\cal P},\forall\,k\in E_{\bf K}\;\;
Agent_P\not\in k,
\ey\right.$}\ee
где ${\cal P}$ -- некоторый РП. 
Напомним, что согласно обозначениям, введенным в конце
пункта \ref{termy}, $E_{\bf K}$ -- это множество термов
из $E$ типа ${\bf K}$ (ключ).

Свойство \re{l6}  истинно
в состоянии $\theta\in\Theta_{\cal P}$
(что обозначается записью
$\theta\models E\,\bot\, P$), если
\be{l7}\!\!\!\!\!\!
\by
\forall\,e\in E,\;
\forall\,e' \in(x_P^\theta)^\theta\cup
x_\circ^\theta
\mbox{ каждое вхождение $e$ в $e'$}\\
\mbox{содержится в подтерме }
k(\ldots)\subseteq e', \mbox{ где } 
 k\in E_{\bf K}.
%
\ey\ee

Данное
свойство имеет следующий
 смысл:
термы из $E$
доступны  процессу $P$ в состоянии $\theta$
только в <<защищённом>> виде,
т.е. содержатся в термах из 
$(x_P^\theta)^\theta\cup
x_\circ^\theta$ только
внутри ШС, которые
зашифрованы на ключах, 
недоступных для $P$ в
  $\theta$.

 \subsection{Сохранение 
  защищённости 
 при переходах }
\label{sadfaaerg}

В этом параграфе  
доказывается теорема
о сохранении свойства защищённости
$E\,\bot\, P$ при переходах РП.
Данная теорема м.б. интерпретирована как 
следующее
утверждение: если 
в текущем состоянии $\theta$
верно свойство
$E\,\bot\, P$,
то никакая собственная
активность процесса  $P$, начиная с состояния $\theta$,
не приведет к тому, 
что какое-либо сообщение
из $E$
когда-нибудь станет
доступным процессу $P$.\\

\refstepcounter{theorem}
{\bf Теорема \arabic{theorem}\label{t1}}

Пусть  задан переход
$\theta\ral{a_{P}}\theta'$
в РП ${\cal P}$.

$\forall\,E\subseteq Public_{\cal P}\cup
tm(Public_{\cal P})_{\bf K}
$ 
верна импликация 
\be{l8}
\theta\models E\,\bot\, P\Rightarrow \theta'\models E\,\bot\, P.\ee

{\bf Доказательство}.


Пусть верна посылка импликации
\re{l8}, т.е. верно
 \re{l7}.
Докажем, что тогда будет верно её заключение,
т.е. 
\be{l9}
\by\forall\,e\in E,\;
\forall\,e' \in(x_P^{\theta'})^{\theta'}\cup x_\circ^{\theta'}
\mbox{ каждое вхождение $e$ в $e'$}\\
\mbox{содержится в подтерме }
k(\tilde u)\subseteq e', \mbox{ где } 
 k\in E_{\bf K}.
\ey\ee

Если  \re{l9}
неверно, то 
\be{l10}
\by
\exists\,e\in E,
\exists\,e'\in (x_P^{\theta'})^{\theta'}\cup x_\circ^{\theta'},
\exists\,\mbox{ вхождение }
e\subseteq e',\mbox{ которое}\\
\mbox{ не содержится в каждом
терме вида }
k(\tilde u)\subseteq e',
\mbox{где }k\in E_{\bf K}.
\ey
\ee

Рассмотрим три варианта перехода
$
\theta\ral{a_{P}}\theta'$.

\bn
\i  $a_P=\circ!u$, 
$u\in tm(x_{P}^\theta),
   x_{P}^{\theta'}=x_{P}^{\theta},
x_\circ^{\theta'}=
x_\circ^{\theta}\cup\{u^\theta\}$.

В этом случае
\re{l10} возможно только 
если $e'=u^\theta$.

Возможен один из следующих двух случаев:
\bn\i $u\in x_P^\theta$,
в этом случае из $e\subseteq u^\theta=e'$
и \re{l7} следует
$$e \subseteq k(\tilde u) \subseteq  
u^\theta = e',\;\mbox{ где } 
k \in  E_{\bf K},$$
что противоречит \re{l10},
\i $u=f(u_1,\ldots, u_n)$, 
где $f\in Fun$, в этом случае противоречивость \re{l10} обосновывается индукцией по структуре $u$: в случае
$e\subset u^\theta$ противоречивость 
\re{l10} следует из индуктивного предположения, 
а в случае $e=u^\theta=e'$,
 согласно свойству \re{l7}, $e$
содержится в подтерме 
$k(\tilde u) \subseteq  e' = e$, т.е.
$e=k(\tilde u)$, что невозможно 
по условию из \re{l6} на множество $E$.

\en

\i $a_P = \circ ?u$, $x_\circ^{\theta'}  = x_\circ^{\theta}$, $x^{\theta'}_P = x^{\theta}_P \cup Var(u)$, 
$u^{\theta'} \in  x_\circ^{\theta}$, 
и если $u^{\theta}$ содержит
подтерм $k(\tilde u)$, где $k \in  Tm_{\bf K}$, то верно 
\re{l4}.

В этом случае \re{l10} возможно только если
\be{l11}\by
\exists\, e \in  E,\; 
\exists\, y \in  Var(u) \setminus  x^{\theta}_P: 
e \subseteq y^{\theta'} = e',\\
\mbox{и не существует терма вида $k(\tilde u)$, где $k \in  E_{\bf K}$},\\
\mbox{такого, что $e \subseteq k(\tilde u) \subseteq 
y^{\theta'}$}.\ey\ee
Поскольку упомянутый в \re{l11} терм $e$ содержится в терме $y^{\theta'} \subseteq u^{\theta'} \in 
x_\circ^{\theta}$, то из \re{l7} следует, что $e$ содержится в некотором подтерме вида
$k(\tilde u) \subseteq u^{\theta'}$, где $k \in  E_{\bf K}$.

Таким образом, $e \subseteq k(\tilde u)$ и $e \subseteq 
y^{\theta'}$, т.е. термы $k(\tilde u)$ и $y^{\theta'}$ имеют
непустое пересечение, поэтому, согласно \re{l2}, 
либо $k(\tilde u) \subseteq y^{\theta'}$, либо
$y^{\theta'} \subset  k(\tilde u)$. Включение $k(\tilde u) \subseteq y^{\theta'}$ противоречит \re{l11}, следовательно
$y^{\theta'} \subset  k(\tilde u)$, поэтому
\be{l12}y^{\theta'}
\subset  k(\tilde u) \subseteq u^{\theta'}. \ee
Докажем индукцией по структуре терма $u$, что из правого включения в \re{l12} следует, что
\be{l13}
\exists\, z \in  Var(u): k(\tilde u) \subseteq z^{\theta'}
\subseteq u^{\theta'}.\ee
Если $u \in  Var$, то $z = u$, если $u \in  Con$, то \re{l12} неверно.

Пусть $u = f(u_1,\ldots, u_n)$, где $f \in  Fun$, тогда возможны следующие
случаи:

\bi
\i  $f = encrypt$, т.е. $u = k_1(u_1)$: если 
$k_1 \in  Var_{\bf K}$, то возможны
следующие случаи:
\bi
\i $k(\tilde u) = u^{\theta'} = k_1^{\theta}
(u_1^{\theta'})$, 
в этом случае $k = k_1^{\theta}
\in  E$, и, согласно
\re{l4}, либо $k \in  Var$, либо
\be{l14}
k = shared\_key(\ldots)\mbox{ и }Agent_P \in  k. \ee

Случай $k \in  Var$ невозможен, т.к. из свойств 
$k \in  (x_P^{\theta})^{\theta}$ и
$k \in  E_{\bf K}$, согласно \re{l7}, 
следует, что вхождение $k$ в $k$ должно
содержаться в подтерме вида $k'(\ldots) \subseteq k$, что невозможно,
а \re{l14} невозможно согласно второй строке в \re{l6},
\i $k(\tilde u) \subseteq k^{\theta'}_1$, данный случай невозможен по определению
термов типа {\bf K},
\i $k(\tilde u) \subseteq u^{\theta'}_1$, в данном случае утверждение \re{l13} следует из
индуктивного предположения,
\ei
а если $k_1 = shared\_key(\ldots)$, то верно \re{l14}, что невозможно
согласно второй строке в \re{l6},
\i  если $f = list$, то $\exists\, i \in  {1, \ldots , n}: k(\tilde u) \subseteq u^{\theta'}_i$, и 
\re{l13} следует
из индуктивного предположения,
\i  случай $f = shared\_key$ невозможен.
\ei
Из \re{l12} и \re{l13} следует, что
\be{l15}y^{\theta'}
\subset  k(\tilde u) \subseteq z^{\theta'}
\subseteq u^{\theta'}. \ee
Таким образом, $u$ содержит вхождения переменных $y$ и $z$, обладающие следующим свойством: $y^{\theta'} \subset  z^{\theta'}$, откуда для данных вхождений
следует включение $y \subset  z$, что невозможно.
\i $a_P = (u := u')$, 
$x_\circ^{\theta'}  = x_\circ^{\theta}$, 
$x^{\theta'}_P = x^{\theta}_P \cup Var(u)$, 
$u' \in  tm(x^{\theta}_P)$, 
$u^{\theta'} = (u')^{\theta}$,
и если $u^{\theta}$ содержит подтерм вида $k(\tilde u)$, 
где $k \in  Tm_{\bf K}$, то верно \re{l4}.

Анализ данного случая аналогичен анализу предыдущего случая.
Свойство \re{l10} верно только если верно свойство 
\re{l11}, из которого
следует, что терм $e$, упомянутый в \re{l11}, обладает свойством $e \subseteq 
u^{\theta'} = (u')^{\theta}$.

Докажем, что
\be{l16}
\exists\, v \in  Var(u') : e \subseteq v^{\theta}. \ee

Т.к. $e \in  E$, то, согласно \re{l6},
\bi
\i  либо $e \in  Var$, в этом случае \re{l16} очевидно,
\i  либо $e$ имеет вид $shared\_key(\ldots)$, в этом случае из $e \subseteq (u')^{\theta'}$,
соотношения \re{l5} и второй строки в \re{l6} следует, что данный случай невозможен.
\ei
Т.к. $u' \in  tm(x^{\theta}_P)$, 
т.е. $Var(u') \subseteq x^{\theta}_P$, то из \re{l16} следует, что $v \in  x^{\theta}_P$,
и $e \subseteq v^{\theta} \in  (x^{\theta}_P )^{\theta}$, откуда, на основании \re{l7}, 
заключаем, что $e \subseteq k(\tilde u) \subseteq v^{\theta} \subseteq (u')^{\theta} = u^{\theta'}$, где 
$k \in  E_{\bf K}$.

Термы $k(\tilde u)$ и $y^{\theta'}$ имеют непустое пересечение (оба содержат $e$), и
из \re{l11} следует, что $k(\tilde u)$ не м.б. подтермом терма $y^{\theta'}$, поэтому из
\re{l2} следует, что
\be{l17}
y^{\theta'}
\subset  k(\tilde u) \subseteq u^{\theta'}.\ee

Так же, как и в предыдущем случае, доказываем, что из 
\re{l17} следует свойство
\be{l18}
\exists\, z \in  Var(u) : k(\tilde u) \subseteq z^{\theta'}
\subseteq u^{\theta'}. \ee

Из \re{l17} и \re{l18} следует, что
\be{l19}y^{\theta'}
\subset  k(\tilde u) \subseteq z^{\theta'}
\subseteq u^{\theta'}. \ee

Таким образом, $u$ содержит вхождения переменных $y$ и $z$, обладающие следующим свойством: $y^{\theta'} \subset  z^{\theta'}$, откуда для данных вхождений
следует включение $y \subset  z$, что невозможно. $\blackbox$
\en

\section{Свойство соответствия}

В этом параграфе формулируется и доказывается теорема, которая может использоваться для обоснования {\bf свойства соответствия} протоколов аутентификации. Данное свойство имеет следующий неформальный
смысл: если один из участников протокола аутентификации (обозначим
его $A$) после выполнения этого протокола пришел к выводу, что другой
участник этого протокола (обозначим его $B$) является подлинным (т.е.
те параметры, которые получил $A$ от якобы участника $B$, совпадают с
теми параметрами, которые $B$ посылал $A$), то $B$ действительно посылал
$A$ сообщение с этими параметрами.

Доказываемая ниже теорема имеет следующий смысл: если при некотором выполнении $\pi$ РП ${\cal P}$ в состоянии 
$\theta \in  \pi$ в канале $\circ$  содержится сообщение, содержащее подтерм $k(e)$, где ключ $k$ недоступен в состоянии $\theta$
для некоторого процесса $P \in  {\cal P}$, то в некотором состоянии 
$\theta' <_{\pi} \theta$ другой
процесс $P'\neq P$ из ${\cal P}$ 
послал в канал $\circ$  сообщение, содержащее $k(e)$.\\

\refstepcounter{theorem}
{\bf Теорема \arabic{theorem}\label{t2}}

Пусть заданы
\bi
\i  РП ${\cal P}$ и некоторое его выполнение 
$\pi = (\theta_0, \theta_1, \ldots)$,
\i  подмножество $E \subseteq  Public_{\cal P} 
\cup tm(Public_{\cal P})_{\bf K}$, и
\i  состояние $\theta \in  \pi$, причем 
$\theta \models  E \bot P$, и $\exists\, e \in  
x_\circ^\theta:$
\be{l20}\mbox{$\exists\, k(\tilde e) \subseteq  e$, где 
$k \in  E_{\bf K}$.} \ee\ei
Тогда $\exists\, P' \in  {\cal P} : P' \neq P$ и $\pi$ содержит переход вида
\be{l21}\dot \theta
\ral{(\circ !\dot e)_{P'}}
\theta',\mbox{ где }k(\tilde e) \subseteq  \dot e^{\dot\theta}\mbox{ и }\theta' \leq_\pi \theta.\ee

{\bf Доказательство}.

Пусть $\theta'$ -- первое состояние на $\pi$, такое, что $\exists\, e' \in  x_\circ^{\theta'}$:
\be{l22}
k(\tilde e) \subseteq  e',\mbox{ где }
k(\tilde e)\mbox{ -- терм из } \re{l20}.\ee

Нетрудно видеть, что $\theta' \leq_\pi \theta$, и т.к. 
$x_\circ^{\theta_0}  = \emptyset$, 
то $\theta' \neq \theta_0$.

Пусть переход на пути $\pi$ с концом в $\theta'$ имеет вид $\dot \theta
\ra{a_{P'}}
\theta'$. Т.к. $e' \not\in  x_\circ^{\dot \theta}$,
то $a_P' = \circ !\dot e$, где $\dot e^{\dot \theta} = e'$.
Если $P'\neq P$, то теорема доказана. 

Докажем,
что случай $P' = P$ невозможен.

Пусть $P' = P$, т.е. $\dot \theta\ral{
(\circ !\dot e)_P} \theta'$.

Если $k = shared\_key(\ldots)$, то, 
согласно \re{l22} и \re{l4}, 
$Agent_P \in  k \in  E_{\bf K}$, что противоречит второй строке в \re{l6}. 
Поэтому
$k \in  Var$.

Из $\theta \models  E \bot P$ следует, что 
$\dot \theta \models  E \bot P$.

Случай $k \in  (x^{\dot \theta}_P )^{\dot \theta}$ невозможен, т.к. из $k \in  (x^{\dot \theta}_P)^{\dot \theta}$ и $k \in  E_{\bf K}$, согласно
\re{l7}, следует, что вхождение $k$ в $k$ должно содержаться в подтерме вида
$k'(\ldots) \subseteq  k$, что невозможно.

Докажем индукцией по структуре терма $\dot e$, что из свойства
\be{l23}
k(\tilde e) \subseteq  \dot e^{\dot \theta}, 
k \in  E_{\bf K}, k \in  Var\mbox{ и }k \not\in  (x^{
\dot \theta}_P )^{
\dot \theta}\ee
следует утверждение
\be{l24}
\exists\, x \in  Var(\dot e) \subseteq  x^{\dot \theta}_P:
 k(\tilde e) \subseteq  x^{\dot \theta}. \ee

Если $\dot e \in  Var$, то $x = \dot e$, а если 
$\dot e \in  Con$, то \re{l23} неверно.

Пусть $\dot e = f(e_1, \ldots , e_n)$, 
где 
$f \in  Fun$, тогда
\bi
\i  если $f = encrypt$, т.е. $\dot e = k_1(e_1)$, 
то возможны следующие случаи:
\bi\i $k(\tilde e) = \dot e ^{\dot \theta} = 
k^{\dot \theta}_1(e^{\dot \theta}_1)$, 
в этом случае
\bi\i $k = k^{\dot \theta}_1$, 
и если $k_1$ имеет вид $shared\_key(\ldots)$, 
то $k = k^{\dot \theta}_1$ тоже
имеет такой вид, но по предположению \re{l23} 
$k \in  Var$,
и
\i если $k_1 \in  Var$, то $k_1 \in  x^{\dot \theta}_P$, 
т.е. $k = k^{\dot \theta}_1 \in  (x^{\dot \theta}_P)^{\dot \theta}$, что противоречит \re{l23},
\ei
\i $k(\tilde e) \subseteq  k^{\dot \theta}_1$, данный случай невозможен по определению термов
типа {\bf K},
\i $k(\tilde e) \subseteq  u^{\dot \theta}_1$, 
в данном случае утверждение \re{l13} 
следует из индуктивного предположения,\ei
\i  если $f = list$, то $\exists\, i \in  {1, \ldots , n}$: $k(\tilde e) \subseteq  e_i^{\dot \theta}$, 
и \re{l13} следует из
индуктивного предположения,
\i  случай $f = shared\_key$ невозможен.\ei

Таким образом, из \re{l24} следует, что 
$\exists\, x \in  x_P^{\dot \theta} : k(\tilde e) 
\subseteq  x^{\dot \theta}$.

Пусть $\theta''$ -- первое состояние на пути $\pi$, 
такое, что $(x^{\theta''}_P )^{\theta''}$ содержит
терм с подтермом $k(\tilde e)$, т.е.
\be{l25}\exists\, x \in  x^{\theta''}_P : 
k(\tilde e) \subseteq  x^{\theta''}.\ee

Из \re{l24} следует, что $\theta'' \leq_\pi \dot \theta$. Нетрудно видеть, что $\theta''$ -- не начальное
состояние, поэтому на пути $\pi$ 
существует ребро вида $\ddot \theta
\ral{a_{P''}}
\theta''$. 
Обозначим $a = a_{P''}$. 
Из определения $\theta''$ следует, что 
$x\not\in  x^{\ddot \theta}_P$, 
поэтому $P'' = P$ (т.к.
при этом переходе изменяется значение $x_P$), и возможны два случая:
\bn
\i $a = \circ ?\ddot e$, $x \in  Var(\ddot e)$, 
$\ddot e^{\theta''} \in  x_\circ^{\ddot \theta}$,\\
т.к. $k(\tilde e) \subseteq  x^{\theta''} \subseteq  
\ddot e^{\theta''} \in  x_\circ^{\ddot \theta}$, 
то получаем противоречие с выбором $\theta'$
как самого первого состояния на пути $\pi$, такого, что 
$x_\circ^{\theta'}$  содержит
терм $e'$ с подтермом $k(\tilde e)$: 
$\ddot \theta$ имеет аналогичное свойство, и находится
левее $\theta'$,
\i $a = (\ddot e := \bar e)$, 
$x \in  Var(\ddot e)$, 
$\bar e \in  tm(x^{\ddot \theta}_P)$, 
$\ddot e^{\theta''}= 
\bar e^{\ddot \theta}$,\\
поскольку
\bi
\i  $k(\tilde e) \subseteq  x^{\theta''} \subseteq  \ddot e^{\theta''} = \bar e^{\ddot \theta}$ и
\i  согласно \re{l23}, $k\not\in (x^{\dot \theta}_P)^{\dot \theta}$, поэтому из $\ddot \theta <_\pi \dot \theta$ следует, что $k\not\in 
(x^{\ddot \theta}_P )^{\ddot \theta}$ 
(т.к. $x^{\ddot \theta}_P \subseteq  x^{\dot \theta}_P$),
\ei
то, аналогично доказательству импликации 
\re{l23} $\Rightarrow$ \re{l24} (заменяя в
ней $\dot e$ на $\bar e$  и $\dot \theta$ на $\ddot \theta$) можно доказать утверждение
$$\exists\, x \in  Var(\bar e) \subseteq  x^{\ddot \theta}_P : k(\tilde e) \subseteq  x^{\ddot \theta},$$
которое противоречит выбору $\theta''$ как самого первого состояния на
пути $\pi$ со свойством \re{l25}: 
$\ddot \theta$ имеет аналогичное свойство, и находится левее $\theta''$. $\blackbox$
\en

\section{Метод верификации протоколов, 
основанный на представленной модели}
\label{s6}

\subsection{Описание метода верификации}

Изложенная в предыдущих пунктах модель криптографических протоколов может применяться для обоснования таких свойств протоколов,
которые представляют собой утверждения следующего типа: если при
каком-либо выполнении $\pi$ анализируемого протокола он достиг некоторого состояния $\theta \in  \pi$, то существуют состояния $\theta', \ldots \leq_\pi \theta$ 
на этом выполнении, 
которые обладают заданными свойствами. В этом пункте в
качестве такого свойства рассматривается свойство соответствия в протоколах аутентификации, определяемое ниже. Метод обоснования этого
свойства заключается в обратном построении выполнения данного протокола, начиная с состояния $\theta$. 
Искомые состояния $\theta', \ldots \leq_\pi \theta$ 
возникают в процессе обратного построения выполнения протокола. Построение
данного выполнения производится с использованием теоремы 
\ref{t2}.

Ниже излагается иллюстрация применения данного метода для верификации свойств соответствия и секретности протокола аутентификации
Yahalom.

\subsection{Описание протокола Yahalom}

Протокол Yahalom предназначен для аутентификации (т.е. проверки подлинности) агентов, взаимодействующих по открытому каналу $\circ$, и передачи сеансовых ключей между этими агентами. Предполагается что
\bi
\i  заданы множество агентов $Ag$, а также агент $J$, называемый {\bf доверенным посредником}, данные агенты могут взаимодействовать
друг с другом по открытому каналу $\circ$,
\i  каждый агент $A \in  Ag$ 
имеет разделяемый ключ $k_{AJ}$ с доверенным
посредником $J$, на котором $A$ и $J$ могут шифровать и расшифровывать
сообщения, используя симметричную систему шифрования.
\ei
В каждом сеансе протокола Yahalom принимают участие следующие
агенты: {\bf инициатор} $A \in  Ag$, 
{\bf доверенный посредник} $J$, и {\bf респондер}
$B \in  Ag$. Каждый агент из $Ag$ в одних сеансах м.б. инициатором, а в
других -- респондером. Выполнение сеанса протокола Yahalom с инициатором $A$, респондером $B$ и доверенным посредником $J$ представляет
собой совокупность четырех пересылок сообщений:
\be{l26}
\by 1.&
A\to B&:&
A,n_A\\2.&
B\to J&:&B, 
k_{BJ}(A, n_A, n_B)\\3.&
J\to A&:&
k_{AJ}(B, k,n_A,n_B), k_{BJ}
(A,
k)\\4.&
A\to B&:&k_{BJ}
(A,
k),k(n_B)
\ey\ee

Пересылки в \re{l26} имеют следующий смысл.
\bn
\i $A$ посылает $B$ запрос
на аутентификацию и генерацию сеансового 
ключа $k$,
этот запрос состоит 
из имени агента $A$ и нонса $n_A$.
\i $B$ посылает $J$ запрос на генерацию 
сеансового ключа $k$,
в свой запрос он включает своё имя, 
имя агента $A$,
для связи с которым нужен этот ключ, 
полученный нонс $n_A$, и свой нонс $n_B$.
\i $J$ генерирует сеансовый ключ $k$
и посылает $A$ пару сообщений, 
из первого сообщения $A$ может извлечь
сеансовый ключ $k$, 
а второе предназначено для
того, чтобы $A$ переслал его 
$B$.
\i $A$ посылает $B$ пару сообщений, 
\bi\i
первое из которых было получено им от $J$, 
$B$ может извлечь из этого сообщения
сеансовый ключ $k$, и
\i используя ключ $k$, 
$B$ расшифровывает второе сообщение. 
\ei
Если 
результат расшифрования совпадает с 
$n_B$, то это является для $B$ 
доказательством того,
что отправителем этого сообщения был
$A$.
\en

\subsection{Некоторые определения и обозначения}

\bn\i
Для каждого процесса $P$ запись $P^*$ обозначает РП
$\{P_i \mid i \in  I\},$
где $I$ -- множество натуральных чисел, и 
$\forall\, i \in  I \;\;P_i = P$.

Будем использовать следующее соглашение:
\bi
\i  если в каком-либо рассуждении, связанном с РП вида 
$P^*$, некоторый процесс является первым из рассматриваемых процессов, входящих в $P^*$, то этот процесс и все его переменные обозначаются теми же записями, которые используются в $P$,
\i  если кроме этого процесса рассматривается другой процесс,
входящий в $P^*$ (возможно совпадающий с $P$), то он обозначается $P_1$, и в обозначениях тех его переменных, которые являются дубликатами переменных из множества
$$Unique_P \cup Private_P \cup \{at_P , x_P \},$$
используется индекс 1 (например, дубликат переменной $x$ в
$P_1$ будет обозначаться записью $x_1$), в следущем процессе ($P_2$,
который возможно совпадает с $P$ или $P_1$) соответствующие
переменные будут обозначаться с индексом 2, и т.д.\ei
\i Процесс называется {\bf линейным}, если он имеет вид
\be{l27}
\by
\begin{picture}(0,0)
\put(-75,0){\oval(8,8)}
\put(-75,0){\circle*{4}}
\put(-25,0){\circle*{4}}
\put(25,0){\circle*{4}}
\put(75,0){\makebox(0,0){$\otimes$}}
\put(75,-7){\makebox(0,0)[t]{$n$}}
\put(25,-7){\makebox(0,0)[t]{$n-1$}}
\put(-25,-7){\makebox(0,0)[t]{$1$}}
\put(-75,-7){\makebox(0,0)[t]{$0$}}
\put(-73,0){\vector(1,0){46}}
\put(27,0){\vector(1,0){44}}
\put(-23,0){\line(1,0){10}}
\put(13,0){\vector(1,0){10}}
\put(0,0){\makebox(0,0){$\ldots$}}
\put(-50,2){\makebox(0,0)[b]{$a_{1}$}}
\put(50,2){\makebox(0,0)[b]{$a_{n}$}}
\end{picture}\ey
\ee

В целях большей наглядности будем использовать следующее соглашение в обозначениях переменных в линейных процессах: пусть
$P$ -- процесс вида \re{l27}, 
и переменная $x$ входит в действие $a_i$, причём
$\forall\, j \in  \{1, \ldots , i - 1\}\;\; x$ не входит в $a_j$, тогда
\bi
\i  если $x \in  Unique_P$, то будем указывать горизонтальную черту над всеми вхождениями $x$ в $a_i$ 
(т.е. обозначать их $\bar x$) и
\i  если $x \in  Private_P$, 
то будем указывать уголок над всеми
вхождениями $x$ в $a_i$ (т.е. обозначать их $\hat x$).\ei

\i Если $\pi$ -- выполнение РП ${\cal P}$, 
то запись $\pi \ni  P^{i,i'}: \theta \ra{a} \theta'$ имеет
следующий смысл: $\pi$ содержит переход $\theta
\ra{a_P}  \theta'$, и $at_P^\theta= i, at_P^{\theta'}
= i'.$
\en

\subsection{Формальное
описание процессов, входящих в
протокол Yahalom}

Описание
сеанса протокола Yahalom изображается
 схемой
$$\by
\begin{picture}(0,20)
\put(-150,0){\oval(8,8)}
\put(-150,0){\circle*{4}}
\put(-80,0){\circle*{4}}
\put(70,0){\circle*{4}}
\put(150,0){\makebox(0,0){$\otimes$}}
\put(-146,0){\vector(1,0){64}}
\put(-78,0){\vector(1,0){146}}
\put(72,0){\vector(1,0){74}}
\put(-180,0){\makebox(0,0){$I_A=$}}
\put(-150,-7){\makebox(0,0)[t]{$0$}}
\put(-80,-7){\makebox(0,0)[t]{$1$}}
\put(70,-7){\makebox(0,0)[t]{$2$}}
\put(150,-7){\makebox(0,0)[t]{$3$}}
\put(-115,2){\makebox(0,0)[b]{$\circ !(A,\bar n^i_A)$}}
\put(-10,2){\makebox(0,0)[b]{$\circ?(k_{AJ}( 
a^r_A,\hat k^i_A,n^i_A,\hat n^r_A), \hat x)$}}
\put(110,2){\makebox(0,0)[b]{$\circ !(x,k^i_A(n^r_A))$}}
\end{picture}\\
\begin{picture}(0,40)
\put(-150,0){\oval(8,8)}
\put(-150,0){\circle*{4}}
\put(-20,0){\circle*{4}}
\put(165,0){\makebox(0,0){$\otimes$}}
\put(-146,0){\vector(1,0){124}}
\put(-18,0){\vector(1,0){179}}
\put(-180,0){\makebox(0,0){$J=$}}
\put(-150,-7){\makebox(0,0)[t]{$0$}}
\put(-20,-7){\makebox(0,0)[t]{$1$}}
\put(165,-7){\makebox(0,0)[t]{$2$}}
\put(-85,2){\makebox(0,0)[b]{$\circ?(\hat
a^r_J, k_{\hat
a^r_JJ}(\hat
a^i_J,\hat n^i_J,
\hat n^r_J)\!)$}}
\put(70,2){\makebox(0,0)[b]{$\circ !(k_{a^i_J J}(a^r_J,\bar k_J,n^i_J,n^r_J), k_{a^r_JJ}
(a^i_J,\bar
k_J))$}}
\end{picture}\\
\begin{picture}(0,40)
\put(-150,0){\oval(8,8)}
\put(-150,0){\circle*{4}}
\put(-85,0){\circle*{4}}
\put(50,0){\circle*{4}}
\put(170,0){\makebox(0,0){$\otimes$}}
\put(-146,0){\vector(1,0){60}}
\put(-84,0){\vector(1,0){132}}
\put(52,0){\vector(1,0){114}}
\put(-180,0){\makebox(0,0){$R_B=$}}
\put(-150,-7){\makebox(0,0)[t]{$0$}}
\put(-85,-7){\makebox(0,0)[t]{$1$}}
\put(50,-7){\makebox(0,0)[t]{$2$}}
\put(170,-7){\makebox(0,0)[t]{$3$}}
\put(-115,2){\makebox(0,0)[b]{$\circ?
(\hat a^i_B, \hat n^i_B)$}}
\put(-20,2){\makebox(0,0)[b]{$\circ !
(B, k_{BJ}(a^i_B,{n^i_B},\bar n^r_B))$}}
\put(105,2){\makebox(0,0)[b]{$\circ?
(k_{BJ}(a^i_B,\hat k^r_B), \hat k^r_B(n^r_B))$}}
\end{picture}\ey$$
\\

В этой схеме
первая и третья диаграммы
 соответствуют 
процессам $I_A$ и $R_B$, 
описывающим поведение инициатора $A$ и респондера $B$ соответственно,
вторая диаграмма соответствует процессу, 
описывающему
поведение посредника $J$, этот процесс 
обозначается символом $J$.
Верхний индекс
$i$ или $r$ при какой-либо переменной
означает, что она 
содержит информацию об 
инициаторе ($i$) или респондере ($r$)
данного сеанса.
Cмысл переменных в этих процессах
усматривается из сопоставления
действий в этих процессах с соответствующими 
действиями в \re{l26}.
Предполагаем, что 
$Agent_{I_A}=A$,
$Agent_{R_B}=B$,
$Agent_J=J.$

РП ${\cal P}$, соответствующий протоколу Yahalom, имеет вид
\be{l28}
\by {\cal P}=
\{
\{I_A^*\mid A\in Ag\}, 
\{R_B^*\mid B\in Ag\}, J^*,\enemy\},\ey\ee
т.е. каждый агент
может участвовать в неограниченном числе сеансов как
в качестве  инициатора, так  и в качестве респондера.

\subsection{Свойства протокола Yahalom}

В этом пункте приводится 
формальное описание и 
верификация трех свойств
протокола
\re{l28}:
секретность ключей
$k_J$
и 
нонсов  $n^r_B$,
аутентификация инициатора перед 
респондером и 
аутентификация респондера перед 
инициатором.
В  доказательствах теорем 
\ref{t3},
\ref{t4},
\ref{t5}
при каждом применении теоремы
\ref{t2}
имеется единственный вариант обоснования
существования перехода \re{l21},
и мы будем 
сразу будем излагать это обоснование,
без упоминания о единственности варианта такого 
обоснования. \\

\refstepcounter{theorem}
{\bf Теорема \arabic{theorem}\label{t3}}
(секретность ключей
$k_J$
и 
нонсов  $n^r_B$)

РП \re{l28}
обладает следующим свойством:
\be{l29}
\forall\, \theta \in  \Theta_{\cal P}\;\; 
\theta \models  E \bot \enemy,\mbox{ где }E = \{k_{BJ} , k_J, n^r_B \mid B \in  Ag\}. \ee

{\bf Доказательство.}

Докажем \re{l29} от противного.

Предположим, что задано выполнение 
$\pi = (\theta_0, \theta_1, \ldots)$, и
$$
S = \{\theta \in  \pi \mid \theta \not \models  E \bot \enemy\}\neq \emptyset.$$

Пусть $\theta$ -- первое состояние на $\pi$, которое принадлежит $S$. 

Т.к. $\theta_0 \models 
E \bot \enemy$, то $\theta \neq \theta_0$, т.е. в $\pi$ есть переход вида $\theta' \ral{a_P} \theta$.

Из определения $\theta$ следует, что 
$\theta' \models  E \bot \enemy$, 
$\theta \not \models  E \bot \enemy$.

Если бы было верно $P = \enemy$, то, согласно теореме \ref{t1},
отсюда следовало бы, что $\theta \models  E \bot \enemy$,
что противоречит определению $\theta$.

Таким образом, $P\in \{I_A,R_B,J\mid 
A,B\in Ag\}$, и нетрудно видеть, что
$a_P$ имеет вид $\circ !e$, 
$x_\circ^\theta  = x_\circ^{\theta'}  \cup \{e^{\theta'}\}$, 
и верно утверждение $\theta \not \models  E \bot \enemy$, которое
в данной ситуации эквивалентно утверждению
\be{l30}\by
\mbox{$\exists\, u \in  E,\; \exists\,$ вхождение 
$u$ в $e^{\theta'}$, не содержащееся}\\
\mbox{ни в каком подтерме вида 
$k(\ldots) \subseteq  e^{\theta'}$, 
где $k \in  E_{\bf K}$.}\ey\ee

Перебором всех возможных обоснований перехода 
$\theta' \ral{\circ !e} \theta$ 
со свойством \re{l30} находим единственное обоснование:
\be{l31}
\pi\ni I_A^{2,3}: \theta'\ra{\circ!e}\theta,\;\;
\mbox{где }e=(x,k^i_A(n^r_A)),\ee
поэтому \re{l30} можно переписать следующим образом:
\be{l32}\by
\mbox{$\exists\, u \in  E, \;\;\exists$ вхождение $u$ 
в $e^{\theta'} = (x, k^i_A
(n^r_A))^{\theta'}$,}\\
\mbox{не содержащееся ни в каком подтерме}\\
\mbox{вида $k(\ldots) \subseteq  (x, k^i_A
(n^r_A))^{\theta'}$, где $k \in  E_{\bf K}$}.
\ey
\ee
Т.к. $\theta' \vdash  at_{I_A} = 2$, то 
$\exists\, \theta_1 \leq_\pi \theta'$:
\be{l33}
\pi\ni
I_A^{1, 2}:
\theta'_1\ral{\circ?e_1}\theta_1,\mbox{ где }
e_1=(k_{AJ}(a^r_A,\hat k^i_A,n^i_A,\hat n^r_A), 
\hat x
).\ee

Т.к. $\theta'_1 <_\pi \theta_1 \leq_\pi \theta'$ и 
$\theta' \models  E \bot \enemy$, то
\be{l34}\theta'_1 \models  E \bot \enemy. \ee
Из \re{l33} следует, что $e_1^{\theta'_1}
 = (\ldots , x^{\theta'_1}) \in  x_\circ^{\theta'_1}$ (многоточия в терме $(\ldots , x^{\theta'_1})$ и
ниже обозначают компоненты, не представляющие интерес для рассмотрения), поэтому, согласно \re{l34} и \re{l7}, верно утверждение:
\be{l35}\by
\mbox{$\forall\, u \in  E$ каждое вхождение $u$ в $x^{\theta'_1} \subseteq  e_1^{\theta'_1}$}\\
\mbox{содержится в подтерме 
$k(\ldots) \subseteq  x^{\theta'_1}$, где 
$k \in  E_{\bf K}$}.
\ey\ee
Сопоставляя \re{l32} и \re{l35}, заключаем:
\be{l36}\by
\mbox{$\exists\, u \in  E, \exists$ вхождение $u$ 
в $(k^i_A
(n^r_A))^{\theta'}$,}\\
\mbox{не содержащееся ни в каком подтерме}\\
\mbox{вида $k(\ldots) \subseteq  (k^i_A
(n^r_A))^{\theta'}$, где $k \in  E_{\bf K}.$}
\ey\ee
По теореме \ref{t2}, из \re{l34}, $e_1^{\theta_1'} \in  
x^{\theta'_1}_
\circ$  и $k_{AJ} \in  E$ следует, что 
$\exists\, \theta_2 \leq_\pi \theta'_1 : \pi$
содержит переход 
$\theta'_2
\ral{(\circ!e_2)_P} 
\theta_2$, 
где $P \in  {\cal P}$ и первая компонента $k_{AJ}(\ldots)$
терма $e_1^{\theta_1'}$ входит в $e_2^{\theta_1'}$. 
Перебором всех вариантов такого перехода находим единственное обоснование:
\be{l37}
\left\{\by\pi\ni
 J^{1, 2}: \theta'_2\ra{\circ!e_2}\theta_2,
\mbox{ где }
e_2=(k_{a^i_{ J} { J}}(a^r_{ J},\bar k_{ J},n^i_{ J},n^r_{ J}), \ldots
)\\
k_{(a^i_{ J})^{\theta} { J}}((a^r_{ J})^{\theta},\bar k_{ J},\ldots
)
=k_{AJ}(a^r_A,(k^i_A)^{\theta},\ldots
)
\ey\right.\ee

Из второй строчки в \re{l37} следует, 
что $(k^i_A)^{\theta} =  k_J$, поэтому из \re{l36}
следует утверждение:
$$\by
\exists\, u \in  E, \exists\, \mbox{ вхождение $u$ в $k_J (\ldots)$, не содержащееся}\\
\mbox{ни в каком подтерме вида $k(\ldots) \subseteq  k_J (\ldots)$, где $k \in  E_{\bf K}$,}\ey$$
которое, очевидно, противоречиво. $\blackbox$\\

\refstepcounter{theorem}
{\bf Теорема \arabic{theorem}\label{t4}}
(аутентификация инициатора перед респондером)

РП \re{l28} обладает следующим свойством: 
$\forall\,R_B\in {\cal P}$, 
$\forall\,\theta\in \Theta_{\cal P}$, 
если
$\theta\vdash at_{R_B}=3$,
то $\exists$ 
$I_A\in {\cal P}$:
\be{l38}
\theta\vdash 
{\def\arraystretch{1}\c{
   at_{I_A}=3\\
a^r_A=B,\; a^i_B=A\\
n^i_A=n^i_B,\;
n^r_A=n^r_B\\k^i_A=k^r_B}}
\ee
{\bf Доказательство.}

Пусть процесс
$R_B\in {\cal P}$ и состояние
$\theta\in \Theta_{\cal P}$
таковы, что
$\theta\vdash at_{R_B}=3$, и $\pi =
(\theta_0, \theta_1, \ldots)$ -- выполнение РП ${\cal P}$, такое, что $\theta \in  \pi$.

Из $\theta\vdash at_{R_B}=3$ следует: $\exists\,\theta_1\leq_{\pi} \theta$:
$$\pi\ni
R_B^{2, 3}:\theta'_1\ra{
\circ?e_1}
\theta_1,\mbox{ где }
e_1=(k_{BJ}(a^i_B,\hat k_B^r), \hat k_B^r(n^r_B)).$$

По теореме \re{t3} верно \re{l29}. Согласно теореме \re{t2}, в этом случае из $e_1^{\theta}
\in
x_\circ^{\theta_1}$
и $k_{BJ}\in E$
следует:
 $\exists\,\theta_2\leq_{\pi} \theta'_1$:
\be{l39}
\left\{
\by
\pi\ni
J^{1, 2}:\theta'_2\ra{\circ!e_2}
\theta_2,\;\;\mbox{где }
e_2=
(\ldots,
k_{a^r_JJ}
(a^i_J,\bar
k_J))\\
k_{(a^r_J)^{\theta}J}
((a^i_J)^{\theta},\bar
k_J)=
k_{BJ}((a^i_B)^{\theta},(k_B^r)^{\theta})
\ey\right.
\ee

Из второй строчки в \re{l39} следует, что
\be{l40}
(a^r_J)^{\theta} = B,(a^i_J)^{\theta}=(a^i_B)^{\theta},
\bar k_J=(k^r_B)^{\theta}.\ee
Из $\theta'_2\vdash at_{J}=1$ 
следует, что 
$\exists\,\theta_3\leq_{\pi} \theta'_2$:
\be{l41}
\pi\ni
J^{0, 1}:\theta'_3\ra{\circ?e_3}
\theta_3,\mbox{ где }
e_3=
(\ldots, 
k_{\hat
a^r_JJ}(\hat
a^i_J,\hat n^i_J,
\hat n^r_J)).\ee

Из \re{l40} и \re{l41} следует, что 
$k_{BJ}(*, *, *)\subseteq
e_3^{\theta}\in x_\circ^{\theta_3}$
(где звёздочки обозначают
некоторые термы),
откуда по теореме
 \re{t2}, с учетом 
соотношений $\theta_3 \models E
\,\bot\, 
{\enemy}$ и $k_{BJ}\in E$
 получаем:
$\exists\,\theta_4\leq_{\pi} \theta'_3$:
\be{l42}
\left\{\by\pi\ni
R_{ B_1}^{1, 2}: \theta'_4\ra{\circ!e_4}\theta_4,
\mbox{ где }e_4=(\ldots, 
k_{{B_1}J}(a^i_{ B_1},{n^i_{ B_1}},\bar n^r_{ B_1}))\\
k_{{ B_1}J}((a^i_{ B_1})^{\theta},(n^i_{ B_1})^{\theta},\bar n^r_{ B_1})=k_{BJ}(
(a^i_B)^{\theta},(n^i_J)^{\theta},
(n^r_J)^{\theta})
\ey\right.
\ee
Из второго равенства в \re{l42} следует, что
\be{l43}
 B_1=B, (n^i_B)^{\theta}=(n^i_J)^{\theta}, \bar n^r_B=(n^r_J)^{\theta}.\ee
По теореме \re{t2}, из 
$\theta_1\models E
\,\bot\, 
{\enemy}$,
$
(k^r_B(n^r_B))^{\theta}
\subseteq e_1^{\theta}
\in
x_\circ^{\theta_1}$, 
и $(k^r_B)^{\theta}=\bar k_J \in E$\\
следует, что
$\exists\,\theta_5\leq_{\pi} \theta'_1$:
\be{l44}
\left\{\by\pi\ni
I_A^{2, 3}: \theta'_5\ra{\circ!e_5}\theta_5,
\mbox{ где }
e_5=(\ldots,
k^i_A(n^r_A))\\
(k^i_A(n^r_A))^{\theta}=\bar k_J(n^r_B)
\ey\right.\ee

Из второй строчки в \re{l44} следует, что
\be{l45}
(k^i_A)^{\theta}=\bar k_J, (n^r_A)^{\theta} = n^r_B.\ee
Из $\theta_5 \vdash at_{I_A}=2$ следует, что 
$\exists\,\theta_6\leq_{\pi} \theta'_5$: 
\be{l46}
\pi\ni
I_A^{1, 2}:
\theta'_6\ra{\circ?e_6}\theta_6,\mbox{ где }
e_6=(k_{AJ}( a^r_A,\hat k^i_A,n^i_A,\hat n^r_A), 
\ldots
).\ee

Из \re{l45} и \re{l46} следует, что
\be{l47}
\by
k_{AJ}(a^r_A,(k^i_A)^{\theta},n^i_A,( n^r_A)^{\theta}) =  k_{AJ}(a^r_A,\bar k_J,n^i_A, n^r_B)
\subseteq e_6^{\theta}\in x_\circ^{\theta_6}.\ey\ee

По теореме \re{t2}, из 
$\theta_6 \models E
\,\bot\, 
{\enemy}$, $k_{AJ}\in E$, и
 \re{l47} следует, что
$\exists\,\theta_7\leq_{\pi} \theta'_6$:

\be{l48}
\left\{\by\pi\ni
 J_1^{1, 2}: \theta'_7\ra{\circ!e_7}\theta_7,
\mbox{ где }
e_7=(k_{a^i_{ J_1} { J_1}}(a^r_{ J_1},\bar k_{ J_1},n^i_{ J_1},n^r_{ J_1}), \ldots
)\\
k_{(a^i_{ J_1})^{\theta} { J}}((a^r_{ J_1})^{\theta},\bar k_{ J_1},(n^i_{ J_1})^{\theta},(n^r_{ J_1})^{\theta})
=k_{AJ}(a^r_A,\bar k_J,n^i_A, n^r_B)
\ey\right.\ee

Из второй строчки в \re{l48} следует, что
\be{l49}
(a^i_{ J_1})^{\theta}=A,
(a^r_{  J_1})^{\theta}=a^r_A,
 J_1=J, 
(n^i_J)^{\theta}=n^i_A, (n^r_J)^{\theta}=
\bar n^r_B.\ee

Свойство \re{l38} следует из \re{l40}, \re{l43}, 
\re{l45}, \re{l49}. $\blackbox$\\

\refstepcounter{theorem}
{\bf Теорема \arabic{theorem}\label{t5}}
(аутентификация респондера перед инициатором)

РП \re{l28} обладает следующим свойством: 
$\forall\, I_A \in  {\cal P}, 
\forall\, \theta \in  \Theta_{\cal P}$, если
$\theta \vdash  at_{I_A} = 2$, то 
$\exists\,R_B \in  {\cal P}$:
\be{l50}
\theta\vdash \c{at_{R_B}=2,\\
a^r_A=B,
a^i_B =A, \\
n^i_A=n^i_B,
n^r_A=n^r_B}.
\ee
{\bf Доказательство.}

Пусть процесс $I_A$ и 
состояние $\theta$ таковы, что 
$\theta \vdash  at_{I_A} = 2$, и 
$\pi =
(\theta_0, \theta_1, \ldots)$ -- выполнение РП 
${\cal P}$, такое, что $\theta \in  \pi$.

Из $\theta \vdash  at_{I_A} = 2$ следует, что 
$\exists\, \theta_1 \leq_\pi \theta$:
\be{l51}
\pi\ni I_A^{1, 2}:
\theta'_1\ra{\circ?e_1}\theta_1,\mbox{ где }
e_1=(k_{AJ}(a^r_A,\hat k^i_A,n^i_A,\hat n^r_A), 
\ldots
).\ee

По теореме \ref{t2},
из $\theta_1\models E
\,\bot\, 
{\enemy}$,
$k_{AJ}(*,*,*,*)
\subseteq e_1^{\theta}
\in
x_\circ^{\theta_1}$
(где звёздочки обозначают
некоторые термы)
и 
$k_{AJ}\in E$,
следует, что 
$\exists\,\theta_2\leq_{\pi} \theta'_1$:

\be{l52}
\left\{
\by\pi\ni
J^{1, 2}:\theta'_2\ra{\circ!e_2}
\theta_2,\mbox{ где }
e_2=
(
k_{a^i_J J}(a^r_J,\bar k_J,n^i_J,n^r_J), \ldots
)\\
k_{(a^i_J)^{\theta} J}((a^r_J)^{\theta},\bar k_J,(n^i_J)^{\theta},(n^r_J)^{\theta})=
k_{AJ}(a^r_A,(k^i_A)^{\theta},
n^i_A,(n^r_A)^{\theta})
\ey\right.\ee

Из второй строчки в \re{l52} следует, что
\be{l53}\by
(a^i_J)^{\theta} =A,
(a^r_J)^{\theta}=a^r_A,
\bar k_J=(k^i_A)^{\theta},\\
(n^i_J)^{\theta}=n^i_A,
(n^r_J)^{\theta}=(n^r_A)^{\theta}.
\ey\ee
Из $\theta_2\vdash at_{J}=1$ 
следует, что 
$\exists\,\theta_3\leq_{\pi} \theta'_2$:
\be{l54}
\pi\ni
J^{0, 1}:\theta'_3\ra{\circ?e_3}
\theta_3,\mbox{ где }
e_3=
(\ldots, 
k_{\hat
a^r_JJ}(\hat
a^i_J,\hat n^i_J,
\hat n^r_J)).\ee
Из \re{l53} и \re{l54} следует, что 
$k_{a^r_AJ}(
A,n^i_A,
(n^r_A)^{\theta}) \subseteq e_3^{\theta}\in x_\circ^{\theta_3}$, откуда
по
теореме \re{t2}, учитывая 
$\theta_3\models E
\,\bot\, 
{\enemy}$,
 и 
$k_{a^r_AJ}\in E$,
получаем:
$\exists\,\theta_4\leq_{\pi} \theta'_3$:
\be{l55}
\left\{\by\pi\ni
R_{ B}^{1, 2}: \theta'_4\ra{\circ!e_4}\theta_4,
\mbox{ где }e_4=(\ldots, 
k_{{B}J}(a^i_{ B},{n^i_{ B}},\bar n^r_{ B}))\\
k_{{ B}J}((a^i_{ B})^{\theta},(n^i_{ B})^{\theta},\bar n^r_{ B})=k_{a^r_AJ}(
A,n^i_A,
(n^r_A)^{\theta}).
\ey\right.
\ee
Второе равенство в \re{l55} влечёт равенства, из которых следует \re{l50}:
$$
B=a^r_A,
(a^i_{ B})^{\theta}=A,
(n^i_{ B})^{\theta}=n^i_A,
\bar n^r_{ B}=(n^r_A)^{\theta}. \;\;\blackbox
$$

\section{Верификация протокола передачи 
сообщений
ме\-жду несколькими агентами}
\label{sdfgstrghw54y46ywh}

В этом пункте  рассматривается 
пример верификации
КП, предназначенного для передачи  ШС
по открытому
каналу 
между несколькими агентами.
Данный КП является обобщением
известного
КП Wide-Mouth Frog и изложен в работе \cite{CM3}.

\subsection{Описание протокола}

Уча\-с\-т\-ники этого протокола -- 
агенты из  множества
 $Ag\subseteq {\it Agents}$ и 
 доверенный посредник $J$.
Каждый агент $A\in Ag$ 
использует для связи с $J$
ключ
$k_{AJ}$, доступный только  $A$ и $J$.
Сеанс передачи  сообщения $x$
в зашифрованном виде
от агента $A\in Ag$ 
агенту $B\in Ag$ 
включает в себя следующие действия:
\bi\i обмен сообщениями между $A$
и $J$, в результате чего 
$J$ 
узнает имя  $A$ отправителя,
имя $B$ получателя, и 
ключ $k$, 
на котором будет зашифровано сообщение $x$
от $A$ для получателя $B$,
\i обмен сообщениями между
$J$ и $B$, в результате чего
$B$ узнает
имя $A$ отправителя 
сообщения, которое $B$ получит от $A$, и 
ключ $k$, на котором будет зашифровано это сообщение,
\i пересылка ШС 
$k(x)$ от $A$ к $B$. 
\ei

Выполнение 
сеанса данного КП с инициатором $A$, 
респондером $B$ и доверенным посредником $J$
представляет собой следующую совокупность 
пересылок
сообщений:
\be{dasfasdfg43rtss}
\by 
1.&A\to J&:& k_{AJ}(A,n_A)\\
2.&J\to A&:& k_{AJ}(n_A,n_J)\\
3.&A\to J&:& k_{AJ}(n_J,k)\\
4.& J\to B&:&k_{BJ}(n_A)\\
5.& B\to J&:&k_{BJ}(n_A, n_B, B)\\
6.& J\to B&:&k_{BJ}(A, n_B, k)\\
7.&A\to B&:&k(x)
\ey\ee

Данный сеанс представляется следующей схемой:
\be{sdfgdsgwer33r333}\hspace{5mm}\by
\begin{picture}(0,152)
\put(-60,152){\makebox(0,0){$I_A$}}
\put(60,152){\makebox(0,0){$R_B$}}
\put(0,152){\makebox(0,0){$J$}}
\put(-60,140){\circle*{4}}
\put(-57,140){\makebox(0,0)[l]{$0$}}
\put(-60,120){\circle*{4}}
\put(-57,120){\makebox(0,0)[l]{$1$}}
\put(-60,100){\circle*{4}}
\put(-57,100){\makebox(0,0)[l]{$2$}}
\put(-60,80){\circle*{4}}
\put(-57,81){\makebox(0,0)[l]{$3$}}
\put(-60,0){\circle*{4}}
\put(-57,0){\makebox(0,0)[l]{$4$}}
\put(0,140){\circle*{4}}
\put(-3,140){\makebox(0,0)[r]{$0$}}
\put(0,120){\circle*{4}}
\put(-3,120){\makebox(0,0)[r]{$1$}}
\put(0,100){\circle*{4}}
\put(-3,100){\makebox(0,0)[r]{$2$}}
\put(-60,130){\vector(1,0){60}}
\put(0,110){\vector(-1,0){60}}
\put(-60,90){\vector(1,0){60}}
\put(-60,10){\vector(1,0){120}}
\put(-60,0){\line(0,1){140}}
\put(-63,130){\makebox(0,0)[r]{$\circ!k_{AJ}(a^r_A,\bar n^i_A)$}}
\put(3,130){\makebox(0,0)[l]{$\circ? k_{\hat a^i_JJ}(\hat a^r_J,\hat n^i_J)$}}
\put(-63,110){\makebox(0,0)[r]{$\circ? k_{AJ}(\bar n^i_A,\hat n^j_A)$}}
\put(3,110){\makebox(0,0)[l]{$\circ! k_{a^i_JJ}(n^i_J,\bar n^j_J)$}}
\put(-63,90){\makebox(0,0)[r]{$\circ! k_{AJ}(n^j_A, \bar k^i_A)$}}
\put(3,92){\makebox(0,0)[l]{$\circ? k_{a^i_JJ}(\bar n^j_J,\hat k_J)$}}
\put(-2,68){\makebox(0,0)[r]{$\circ! k_{a^r_JJ}(n^i_J)$}}
\put(63,70){\makebox(0,0)[l]{$\circ? k_{BJ}(\hat n^i_B)$}}
\put(-3,50){\makebox(0,0)[r]{$\circ? k_{a^r_JJ}(n^i_J, \hat n^r_J,a^r_J)$}}
\put(63,50){\makebox(0,0)[l]{$\circ! k_{BJ}(n^i_B,\bar n^r_B,B)$}}
\put(-3,30){\makebox(0,0)[r]{$\circ! k_{a^r_JJ}(a^i_J,n^r_J,
 k_J)$}}
\put(63,30){\makebox(0,0)[l]{$\circ? k_{BJ}(
\hat a^i_B,\bar  n^r_B,\hat k^i_B)$}}
\put(-63,10){\makebox(0,0)[r]{$
\circ!\bar k^i_{A}( x^i_A)$}}
\put(63,10){\makebox(0,0)[l]{$\circ?k^i_B(\hat x^i_B)$}}

\put(0,80){\circle*{4}}
\put(3,79){\makebox(0,0)[l]{$3$}}
\put(0,60){\circle*{4}}
\put(3,60){\makebox(0,0)[l]{$4$}}
\put(0,40){\circle*{4}}
\put(3,40){\makebox(0,0)[l]{$5$}}
\put(0,20){\circle*{4}}
\put(3,20){\makebox(0,0)[l]{$6$}}
\put(60,80){\circle*{4}}
\put(57,79){\makebox(0,0)[r]{$0$}}
\put(60,60){\circle*{4}}
\put(57,60){\makebox(0,0)[r]{$1$}}
\put(60,40){\circle*{4}}
\put(57,40){\makebox(0,0)[r]{$2$}}
\put(60,20){\circle*{4}}
\put(57,20){\makebox(0,0)[r]{$3$}}
\put(60,0){\circle*{4}}
\put(57,0){\makebox(0,0)[r]{$4$}}
\put(65,-5){\makebox(0,0)[l]{$P_B$}}
\put(0,70){\vector(1,0){60}}
\put(60,50){\vector(-1,0){60}}
\put(0,30){\vector(1,0){60}}
\put(60,0){\line(0,1){80}}
\put(0,20){\line(0,1){120}}
\end{picture}\ey
\ee

РП ${\cal P}$, соответствующий этому
КП, имеет вид
\be{sdgffdsghdsfe334w}
\by {\cal P}=
\{
\{I_A^*\mid A\in Ag\}, 
\{R_B^*\mid B\in Ag\}, J^*,\enemy\}.\ey\ee

Свойства этого КП, которые должны
быть верифицированы:
\bi
\i {\bf секретность}
ключей, 
передаваемых сообщений и нонсов:
\be{dfgdsf3241325h3g}
\forall\,c\in \Theta_{{\cal P} }\;\;
\theta\models E
\,\bot\,
\enemy ,\;\mbox{ где }
E=\{k_{AJ}, k^i_A, x^i_A, n^i_A\mid A\in Ag\}
\ee
\i {\bf целостность} передаваемых сообщений:
\be{sdfgad32sgfd2sgsdf332ds}\by
\mbox{$\forall\,R_B\in {\cal P}$, 
$\forall\,\theta\in \Theta_{{\cal P} }$, 
если
$\theta\vdash at_{R_B}=4$,
то $\exists$ 
$I_A\in {\cal P}$:
}
\\
\theta\vdash \{at_{I_A}=4,
a^r_A=B, a^i_B=A, 
 n^i_A=n^i_B,
k^i_A=k^i_B,
x^i_A=x^i_B
\}\ey\ee
\ei

\subsection{Верификация протокола}
\label{fgsrh654sey5tg}

Доказательство свойства
секретности \re{dfgdsf3241325h3g}
дословно повторяет начало рассуждений
по доказательству аналогичного свойства 
протокола Yahalom.

Докажем свойство целостности
\re{sdfgad32sgfd2sgsdf332ds}.
Будем использовать в этом доказательстве
свойство 
\re{dfgdsf3241325h3g}
(не упоминая об этом).

Пусть процесс
$R_B\in {\cal P}$ и состояние
$\theta\in \Theta_{{\cal P} }$
таковы, что
$\theta\vdash at_{R_B}=4$.
Докажем, что 
$\exists\,I_A\in  {\cal P}$: 
выполнено утверждение во второй строчке 
\re{sdfgad32sgfd2sgsdf332ds}.

Пусть $\pi$ -- путь из $\theta^0$ в $\theta$.
Из $\theta\vdash at_{R_B}=4$ следует, что 
\be{zdfgdfgdsghsrth6srje34}
\by
\exists\,\theta_1\leq_{\pi} \theta: \;
\pi\ni
R_B^{3, 4}:\theta'_1\ra{
\circ?e_1}
\theta_1,\mbox{ где }
e_1=k^i_B(\hat x^i_B)
\\
\exists\,\theta_2\leq_{\pi} \theta'_1: \;
\pi\ni
R_B^{2, 3}:\theta'_2\ra{
\circ?e_2}
\theta_2,\mbox{ где }
e_2=k_{BJ}(\hat a^i_B, \bar n^r_B,\hat k^i_B)
\ey\ee

По теореме 2,
из второй строки в \re{zdfgdfgdsghsrth6srje34},
$e_2^{\theta}\in x_\circ^{\theta_2}$,
$k_{BJ}\in E$,
следует:
\be{fgf32dsgdsfg3dsfg4dsfgdsfgds4rhsg}
\left\{
\by\exists\,\theta_3\leq_{\pi} \theta'_2:
\pi\ni
J^{5, 6}:\theta'_3\ra{\circ!e_3}
\theta_3,\mbox{ где }
e_3=k_{a^r_JJ}(a^i_J,n^r_J,
 k_J)\\
k_{(a^r_J)^\theta J}((a^i_J)^\theta,(n^r_J)^\theta,
 (k_J)^\theta)=
k_{BJ}(\hat a^i_B, \bar n^r_B,\hat k^i_B)
\ey\right.\ee

Из второй строки 
в \re{fgf32dsgdsfg3dsfg4dsfgdsfgds4rhsg}
следует, что
\be{afdgsr4ew56yh5eagre}
(a^r_J)^\theta = B,\;
(a^i_J)^\theta=(a^i_B)^\theta,\;
(n^r_J)^\theta=\bar n^r_B,\;
(k_J)^\theta= (k^i_B)^\theta.
\ee

Из первой строки 
в \re{fgf32dsgdsfg3dsfg4dsfgdsfgds4rhsg},
с учетом \re{afdgsr4ew56yh5eagre},
получаем:
\be{arfgste3w5wty6}
\exists\,\theta_4\leq_{\pi} \theta'_3:
\pi\ni
J^{4, 5}:\theta'_4\ra{\circ?e_4}
\theta_4,\mbox{ где }
e_4=k_{BJ}(n^i_J, n^r_B,B).\ee

По теореме 2,
из \re{arfgste3w5wty6},
и того, что $e_4^{\theta}\in x_\circ^{\theta_4}$,
$k_{BJ}\in E$,
следует:
\be{21fgf32dsgdsfg433dsfg4dsfgdsfgds4rhsg}
\left\{
\by\exists\,\theta_5\leq_{\pi} \theta'_4:
\pi\ni
 B_1^{1, 2}:\theta'_5\ra{\circ!e_5}
\theta_5,\mbox{ где }
e_5=k_{ B_1J}(n^i_{ B_1},\bar n^r_{ B_1},
{ B_1})
\\
k_{{ B_1}J}((n^i_{{ B_1}})^\theta,
\bar n^r_{{ B_1}},
{{ B_1}})
=
k_{BJ}((n^i_J)^\theta, \bar n^r_B,B)
\ey\right.\ee

Из второй строки 
в \re{21fgf32dsgdsfg433dsfg4dsfgdsfgds4rhsg}
следует, что 
\be{fdsgsr546uh76yrhtg}
\bar n^r_{ B_1}= \bar n^r_{B},\; 
 B_1 = B, \;
(n^i_{{ B}})^\theta=(n^i_J)^\theta.\;
\ee 
%


Из \re{arfgste3w5wty6}
следует, что
\be{arfgste3w5wt443y6}
\exists\,\theta_6\leq_{\pi} \theta'_4:
\pi\ni
 J^{2, 3}:\theta'_6\ra{\circ?e_6}
\theta_6,\mbox{ где }
e_6=k_{a^i_{ J}J}(n^j_{ J},  k_{ J}).\ee

По теореме 2,
из \re{arfgste3w5wt443y6},
и того, что $e_6^{\theta}\in x_\circ^{\theta_6}$,
$k_{(a^i_{ J})^\theta J}\in E$,
следует:
\be{2132fg4f3244dsfgdsfgds4rhsg}
\left\{
\by\exists\,\theta_7\leq_{\pi} \theta'_6:
\pi\ni
A^{2, 3}:\theta'_7\ra{\circ!e_7}
\theta_7,\mbox{ где }
e_7=k_{AJ}(n^j_A, \bar k^i_A)
\\
k_{AJ}((n^j_A)^\theta, \bar k^i_A)
=
k_{(a^i_{ J})^\theta J}(\bar n^j_{ J},  (k_{ J})^\theta)
\ey\right.\ee

Из второй строки 
в \re{2132fg4f3244dsfgdsfgds4rhsg}
получаем:
\be{fdgdsfdsgrege1}
A=(a^i_{ J})^\theta,\;
(n^j_A)^\theta=\bar n^j_{ J},\;
\bar k^i_A=(k_{ J})^\theta
\ee

Из первой строки 
в \re{2132fg4f3244dsfgdsfgds4rhsg}
получаем:
\be{arfgste3w325wt443y6}
\exists\,\theta_{8}\leq_{\pi} \theta'_7:
\pi\ni
I_A^{1, 2}:\theta'_{8}\ra{\circ?e_{8}}
\theta_{8},\mbox{ где }
e_{8}=k_{AJ}(\bar n^i_A,\hat n^j_A).\ee

По теореме 2,
из \re{arfgste3w325wt443y6},
и того, что $e_{8}^{\theta}\in x_\circ^{\theta_{8}}$,
$k_{AJ}\in E$,
следует:
\be{2132fg4f324324dsfgdsfgds4rhsg}
\left\{
\by\exists\,\theta_{9}\leq_{\pi} \theta'_{8}:
\pi\ni
{ J_1}^{1, 2}:\theta'_{9}\ra{\circ!e_{9}}
\theta_{9},\mbox{ где }
e_{9}=k_{a^i_{{ J_1}} J}(n^i_{{ J_1}}, \bar n^j_{{ J_1}})
\\
k_{(a^i_{{ J_1}})^\theta J}((n^i_{{ J_1}})^\theta, \bar n^j_{{ J_1}})
=
k_{AJ}(\bar n^i_A,(n^j_A)^\theta)
\ey\right.\ee

Из второй строки 
в \re{2132fg4f324324dsfgdsfgds4rhsg}
получаем:
\be{fdgdsfdsgrege2}
(a^i_{{ J_1}})^\theta =A,\;
(n^i_{{ J_1}})^\theta=\bar n^i_A,\;
\bar n^j_{{ J_1}}=
(n^j_A)^\theta
\ee

Из \re{fdgdsfdsgrege1} и
\re{fdgdsfdsgrege2}
получаем: 
\be{hrdttrhtrshstrdryhtrse}
\bar n^j_{ J_1}=\bar n^j_{{ J}}=(n^j_A)^\theta,\;
 {{ J_1}} = { J},\;
 (a^i_{{ J}})^\theta =A,\;
(n^i_{{J}})^\theta=\bar n^i_A.
\ee

Из \re{afdgsr4ew56yh5eagre}
и  \re{fdgdsfdsgrege1}
получаем:
\be{dfgdsgw45g46h4y}
(k^i_B)^\theta =(k_J)^\theta= \bar k^i_A\in E,\ee
поэтому
по теореме 2,
из первой строки в 
\re{zdfgdfgdsghsrth6srje34}
и $e_{1}^{\theta}\in x_\circ^{\theta_{1}}$
следует:
\be{23232132fg4f324324dsfgdsfgds4rhsg}
\left\{
\by\exists\,\theta_{11}\leq_{\pi} \theta'_{1}:
\pi\ni
{ A_1}^{3,4}:\theta'_{11}\ra{\circ!e_{11}}
\theta_{11},\mbox{ где }
e_{11}=\bar k^i_{ A_1}(x^i_{ A_1})
\\
\bar k^i_{ A_1}(x^i_{ A_1})=
(k^i_B)^\theta((x^i_B)^\theta)
\ey\right.\ee

Из второй строки в 
\re{23232132fg4f324324dsfgdsfgds4rhsg}
и 
\re{dfgdsgw45g46h4y}
получаем:
\be{sgfasdfaga43w35hyafgd}
\bar k^i_{ A_1}=(k^i_B)^\theta=\bar k^i_A,\;
 A_1=A, \;
x^i_{A}=(x^i_B)^\theta.
\ee

Из первой строки 
в \re{2132fg4f324324dsfgdsfgds4rhsg}
и \re{hrdttrhtrshstrdryhtrse}
получаем:
\be{hdsfjldskj53438w}
\exists\,\theta_{10}\leq_{\pi} \theta'_{9}:
\pi\ni
 J^{0, 1}:\theta'_{10}\ra{\circ?e_{10}}
\theta_{10},\mbox{ где }
e_{10}=k_{\hat a^i_{ J}J}(\hat a^r_{ J},
\hat n^i_{ J}).\ee

Из \re{afdgsr4ew56yh5eagre} и \re{hrdttrhtrshstrdryhtrse}
следует, что
$e^\theta_{10}=k_{AJ}(B, \bar n^i_A)$.

По теореме 2,
из \re{hdsfjldskj53438w},
 $e_{10}^{\theta}\in x_\circ^{\theta_{10}}$,
$k_{AJ}\in E$,
следует:
\be{213322fg4f324324dsfgdsfgds4rhsg}
\left\{
\by\exists\,\theta_{12}\leq_{\pi} \theta'_{10}:
\pi\ni
{ A_1}^{0, 1}:\theta'_{12}\ra{\circ!e_{12}}
\theta_{12},\mbox{ где }
e_{12}=k_{{ A_1} J}(a^r_{{ A_1}}, \bar n^i_{{ A_1}})
\\
k_{ A_1 J}(a^r_{ A_1},\bar n^i_{ A_1})=
k_{AJ}(B, \bar n^i_A)
\ey\right.\ee

Из второй строки 
в \re{213322fg4f324324dsfgdsfgds4rhsg}
получаем:
\be{fd32gdsfdsgrege2}
\bar n^i_{{ A_1}}=\bar n^i_A,\;
 A_1 = A,\;
a^r_{{ A}} = B.
\ee

Утверждение 
\re{sdfgad32sgfd2sgsdf332ds}
обосновывается следующим образом:
\bi
\i $\theta\vdash at_{I_A}=4$ следует из 
\re{23232132fg4f324324dsfgdsfgds4rhsg},
\re{sgfasdfaga43w35hyafgd}:
$\theta_{11}\vdash at_{ A_1}=4$,
$ A_1=A$,  $\theta_{11}\leq_\pi s$,
\i $\theta\vdash a^r_A=B$ следует из
\re{fd32gdsfdsgrege2},
\i $\theta\vdash a^i_B=A$ следует из 
\re{afdgsr4ew56yh5eagre} и \re{fdgdsfdsgrege1},
\i $\theta\vdash n^i_A=n^i_B$ следует из
\re{fdsgsr546uh76yrhtg} и
\re{hrdttrhtrshstrdryhtrse},
\i $\theta\vdash k^i_A=k^i_B$ следует из
\re{dfgdsgw45g46h4y},
\i $\theta\vdash x^i_A=x^i_B$ следует из
\re{sgfasdfaga43w35hyafgd}. $\blackbox$
\ei

\section{Заключение}

В настоящей работе была построена новая модель КП, и показаны примеры ее использования для решения задач верификации свойств  секретности и соответствия.

Для дальнейшей деятельности
по развитию данной модели и основанных на ней методов
верификации можно назвать следующие
задачи:
\bi
\i развитие языков спецификаций свойств КП, позволяющих выражать например 
свойства нулевого разглашения в КП аутентификации, 
свойства неотслеживаемости в КП электронных платежей,
свойства анонимности и правильности подсчета голосов в КП
электронного голосования,
и разработка методов верификации свойств,
выражаемых на этих языках,
\i построение методов автоматизированного синтеза
КП по описанию свойств, 
которым они должны удовлетворять.
\ei

\end{document}

\bibitem
{n1}
Veronique Cortier, Stephanie Delaune, and Vaishnavi Sundararajan. 
A Decidable Class of Security Protocols for Both Reachability and Equivalence Properties. Journal of Automated Reasoning, 65:479–520, April 2021.

\bibitem
{n324}
Roggenbach, M., Cerone, A., Schlingloff, H., Schneider, G., Shaikh, S.A.,
Formal verification of security
protocols, in:
Formal Methods for Software Engineering: Languages, Methods, Application Domains (Texts in Theoretical Computer Science. An EATCS Series) 1st ed.,
Springer International Publishing, 
2021.

\bibitem
{n2}
Veronique Cortier and Cyrille Wiedling. 
A formal analysis of the Norwegian E-voting protocol. Journal of Computer Security, 25(15777):21–57, 2017.

\bibitem
{mypapersabf}
M. Abadi, B. Blanchet, C. Fournet. The Applied Pi Calculus: Mobile Values, New
Names, and Secure Communication. [Research Report] ArXiv. 2016, pp.110. \verb'hal-01423924',
\verb'https://arxiv.org/abs/1609.03003'

\bibitem
{mypap32ersabf}
Bruno Blanchet, Modeling and Verifying Security Protocols with the Applied Pi Calculus and ProVerif, 2016.

\bibitem
{mypapersae2bf}
Fan Yang,
Santiago Escobar,
Catherine A Meadows,
Jose Meseguer.
Strand Spaces with Choice via a Process Algebra Semantics.
PPDP '16: Proceedings of the 18th International Symposium on Principles and Practice of Declarative Programming, September 2016, pages 76–89.


\bibitem
{mypapersverc}
Veronique Cortier, Steve Kremer.
Formal Models and Techniques for Analyzing Security Protocols: A Tutorial. Foundations and Trends in Programming Languages, 1(3):151–267, (2014)

\bibitem
{myre3papersverc}
Yongjian Li, Jun Pang.
An inductive approach to strand spaces.
Formal Aspects of Computing, Vol. 25, No. 4, 2013.

\bibitem
{mypa23persstrandlast}
Cas Cremers, Sjouke Mauw.
Operational Semantics and Verification of Security Protocols, 
Springer-Verlag Berlin Heidelberg,
2012.

\bibitem
{mypapersstrandlast}
Joshua D. Guttman.
 State and Progress in Strand Spaces: Proving Fair Exchange. Journal of Automated Reasoning, 48(2): 159-195, 2012. 
 
\bibitem
{C324M6}
V. Cortier and S. Kremer, editors. Formal Models and Techniques for Analyzing
Security Protocols, volume 5 of Cryptology and Information Security
Series. IOS Press, 2011.

\bibitem
{CM67} A. Datta, J.C. Mitchell, A. Roy, S. Stiller, Protocol composition logic, in Formal Models
and Techniques for Analyzing Security Protocols, ed. by V. Cortier, S. Kremer (IOS Press,
Lansdale, 2011)

\bibitem
{mypapersrsm}
Mark D. Ryan and Ben Smyth, Applied pi calculus, 
in: Formal Models and Techniques for Analyzing Security Protocols, Edited by Veronique Cortier,  2011 IOS Press, p. 112-142.

\bibitem
{CM54} C.J.F. Cremers, On the protocol composition logic PCL, in ACM Symposium on Information, Computer \& Communication Security (ASIACCS’08), ed. by M. Abe, V. Gligor, Tokyo,
Japan (ACM, New York, 2008), pp. 66--76

\bibitem
{mypaperskerberos}      
Cervesato I., Jaggard A.D., Scedrov A., Tsay J.-K., Walstad C.,
Breaking and fixing public-key Kerberos,
Information and Computation
Volume 206, Issues 2-4, (2008), Pages 402-424.

\bibitem
{CM1} M. Abadi, B. Blanchet, C. Fournet, Just Fast Keying in the Pi Calculus. 
In ACM Transactions on Information and System Security, 10(3), 2007.

\bibitem
{CM66} A. Datta, A. Derek, J.C. Mitchell, A. Roy, Protocol Composition Logic (PCL), in Computation, Meaning, and Logic: Articles dedicated to Gordon Plotkin, ed. by L. Cardelli, M. Fiore,
G. Winskel. Electronic Notes in Theoretical Computer Science, vol. 172, (2007), pp. 311--
358

\bibitem
{CM72} S. Doghmi, J.D. Guttman, F.J. Thayer, Skeletons and the shapes of bundles, in 7th International Workshop on Issues in the Theory of Security (WITS’07), Braga, Portugal (2007)

\bibitem
{CM74} S.F. Doghmi, J.D. Guttman, F.J. Thayer, Skeletons, homomorphisms, and shapes: characterizing protocol executions, in 23rd Conference on the Mathematical Foundations of Programming Semantics (MFPS XXIII), New Orleans, USA. Electronic Notes in Theoretical
Computer Science, vol. 173 (Elsevier, Amsterdam, 2007), pp. 85--102

\bibitem
{CM73} S.F. Doghmi, J.D. Guttman, F.J. Thayer, Searching for shapes in cryptographic protocols, in
13th International Conference on Tools and Algorithms for the Construction and Analysis
of Systems (TACAS’07), ed. by O. Grumberg, M. Huth, Braga, Portugal. Lecture Notes in
Computer Science, vol. 4424 (Springer, Berlin, 2007), pp. 523--537

\bibitem
{mypapershartog3}
M. Abadi and B. Blanchet. 
Analyzing Security Protocols with Secrecy Types and Logic Programs. 
In Journal of the ACM, 52(1), pp. 102-146, 2005.

\bibitem
{32CM21}
I. Cervesato,
N. Durgin,
P. Lincoln,
J. Mitchell,
A. Scedrov.
A Comparison between Strand Spaces and Multiset Rewriting for Security Protocol Analysis.
Journal of Computer Security, vol. 13, no. 2, pp. 265-316, 2005

\bibitem
{mypapershartog10}
S. Kremer, M. Ryan. 
Analysis of an Electronic Voting Protocol in the Applied Pi Calculus. 
In 14th European Symposium on Programming (ESOP), pp. 186-200, 2005.

\bibitem
{CM91} J.D. Guttman, F.J. Thayer, Authentication tests and the structure of bundles. Theor. Comput.
Sci. 283(2), 333--380 (2002)

\bibitem
{CM151} S.G. Stubblebine, R.N. Wright, An authentication logic with formal semantics supporting
synchronization, revocation, and recency. IEEE Trans. Softw. Eng. 28(3), 256--285 (2002)

\bibitem
{CM2} M. Abadi, C. Fournet, Mobile values, new names, and secure communication, in 28th ACM
SIGPLAN-SIGACT Symposium on Principles of Programming Languages (POPL’01), ed. by
C. Hankin, D. Schmidt, London, UK (ACM, New York, 2001), pp. 104--115

\bibitem
{mypapershartog4}
B. Blanchet. 
An Efficient Cryptographic Protocol Verifier Based on Prolog Rules. In 14th IEEE Computer Security Foundations Workshop (CSFW), pp. 82-96, 2001.

\bibitem
{CM78} N.A. Durgin, J.C. Mitchell, D. Pavlovic, A compositional logic for protocol correctness,
in 14th IEEE Computer Security Foundations Workshop (CSFW’01), Cape Breton, Canada
(IEEE Computer Society, Los Alamitos, 2001), pp. 241--272


\bibitem
{8} G. Bella. Inductive Verification of Cryptographic Protocols. PhD thesis,
Cambridge University, 2000.

\bibitem
{38} J. D. Guttman and F. J. Thayer. Authentication tests and the normal,
efficient penetrator. IEEE Computer Society Symposium on Research in Security
and Privacy, 2000.

\bibitem
{CM1321}
P.Y.A. Ryan, S.A. Schneider,
M.H. Goldsmith, G. Lowe and A.W. Roscoe.
The Modelling and Analysis of Security Protocols: the CSP Approach,
Addison-Wesley, 2000.

\bibitem
{82} P. Y. A. Ryan and S. A. Schneider. Process algebra and non-interference. Journal
of Computer Security, 2000. 

\bibitem
{CM3} M. Abadi, A.D. Gordon, A calculus for cryptographic protocols: the Spi calculus. Inf. Comput. 148, 1--70 (1999)

\bibitem
{mypapershartog5}
L. C. Paulson. 
Inductive Analysis of the Internet Protocol TLS. 
In ACM Trans. on Information and System Security, 2(3), pp. 332-351, 1999.

\bibitem
{mypapersstrand1}
F. J. Thayer, J. C. Herzog, and J. D. Guttman. 
Strand spaces: Proving security protocols correct. 
Journal of Computer Security, 7(2/3):191-230, 1999. 

\bibitem
{CM154} F.J. Thayer, J.C. Herzog, J.D. Guttman, Mixed Strand Spaces, in 12th IEEE Computer Security Foundations Workshop (CSFW’99),  IEEE Computer Society, Los
Alamitos, 1999, pp. 72--82

\bibitem
{CM131} L.C. Paulson, The inductive approach to verifying cryptographic protocols. J. Comput. Secur.
6(1--2), 85--128 (1998)

\bibitem
{CM153} F.J. Thayer, J.C. Herzog, J.D. Guttman, Honest ideals on Strand Spaces, in 11th IEEE Computer Security Foundations Workshop (CSFW’98), Rockport, USA (IEEE Computer Society,
Los Alamitos, 1998), pp. 66--77

\bibitem
{30} F. J. Thayer, J. C. Herzog, and J. D. Guttman. Strand spaces: why is a
security protocol correct? IEEE Computer Society Symposium on Security and
Privacy, 1998.

\bibitem
{86} S. A. Schneider. Verifying authentication protocols in CSP. IEEE Transactions
on Software Engineering, 1998.

\bibitem
{27} B. Dutertre and S. A. Schneider. Embedding CSP in PVS. An application to
authentication protocols. Theorem proving in Higher Order Logics, number 1275
in LNCS. Springer, 1997.

\bibitem
{61} G. Lowe and A. W. Roscoe. Using CSP to detect errors in the TMN protocol.
IEEE Transactions in Software Engineering, 23(10), 1997.

\bibitem
{CM130} L.C. Paulson, Proving properties of security protocols by induction, in 10th IEEE Computer
Security Foundations Workshop (CSFW’97), Rockport, Massachusetts (IEEE Computer Society, Los Alamitos, 1997), pp. 70--83

\bibitem
{CM144} S. Schneider, Security properties and CSP, in 17th IEEE Symposium on Security \& Privacy
(S\&P’96), Oakland, USA (IEEE Computer Society, Los Alamitos, 1996), pp. 174--187.

\bibitem
{CM152} P.F. Syverson, P.C. van Oorschot, A unified cryptographic protocol logic. CHACS Report
5540-227 NRL (1996)

\bibitem
{88} S. A. Schneider and A. Sidiropoulos. CSP and anonymity. European Symposium
on Research in Computer Security, 1996.

\bibitem
{5} R. Anderson and R. Needham. Programming Satan's computer. In J. van Leeuwen
(ed.) Computer Science Today, volume 1000 of LNCS. Springer, 1995.

\bibitem
{r2332}
Gavin Lowe. An attack on the Needham-Schroeder public key authentication protocol. Information Processing Letters, 56(3):131--136, November 1995.

\bibitem
{CM99} R.A. Kemmerer, C. Meadows, J.K. Millen, Three systems for cryptographic protocol analysis. J. Cryptol. 7, 79--130 (1994)

\bibitem
{CM126} P.C. van Oorschot, Extending cryptographic logics of belief to key agreement protocols,
in 1st ACM Conference on Computer and Communications Security (ACM CCS’93), ed. by
D.E. Denning, R. Pyle, R. Ganesan, R.S. Sandhu, V. Ashby, Fairfax, USA (ACM, New York,
1993), pp. 232--243

\bibitem
{mypapers11} Syverson P., Meadows C., A Logical Language for Specifying
     Cryptographic Protocol Requirements, Proceedings of the 1993 IEEE
	 Computer Security Symposium on Security and Privacy, (1993) 165-177,
     IEEE Computer Society Press.

\bibitem
{CM5} M. Abadi, M. Tuttle, A semantics for a logic of authentication, in 10th ACM Symposium
on Principles of Distributed Computing (PODC’91), Montreal, Canada (ACM, New York,
1991), pp. 201--216

\bibitem
{mypapers5} Burrows M., Abadi M., Needham R., 
A Logic of Authentication. In ACM
     Transactions on Computer Systems, 8(1), (1990) 18-36.

\bibitem
{CM87} L. Gong, R.M. Needham, R. Yahalom, Reasoning about belief in cryptographic protocol
analysis, in 11th IEEE Symposium on Security \& Privacy (S\&P’90), Oakland, USA (IEEE
Computer Society, Los Alamitos, 1990), pp. 234--248

\bibitem
{mypapers3} Needham R., Schroeder M., Authentication revisited, Operating Systems
     Review, Vol. 21, No. 1, (1987).

\bibitem
{41} C. A. R. Hoare. Communicating Sequential Processes. Prentice-Hall, 1985.

\bibitem
{mypapers1}
 Denning D., Sacco G., Timestamps in Key Distribution Protocols,
     Communications of the ACM, Vol. 24, No. 8, (1981) 533-536.
     
\bibitem
{milner}
R. Milner, A Calculus of Communicating Systems, Springer Verlag, 1980.

\bibitem
{needhams}
Roger Needham and Michael Schroeder. Using encryption for authentification in large networks of computers. Communications of the ACM, 21(12), December 1978.

\bibitem
{afpi} M. Abadi and C. Fournet. Mobile values, new names, and secure communication. In POPL'01: Proceedings of the 28th ACM SIGPLAN-SIGACT Symposium on
Principles of Programming Languages, pages 104–115. ACM Press, 2001.

\bibitem
{apical}
M. Abadi, B. Blanchet, and C. Fournet.
2017. The Applied Pi Calculus: Mobile Values, New Names,
and Secure Communication. J. ACM 65, 1, Article 1 (October 2017), 103 pages.

\bibitem
{modelchec}
Миронов А.М.
Методы верификации программ.
ДМК Пресс Москва, 2023, 335 с.

\end{thebibliography}

\end{document}

\section{}

\begin{center}
{\bf Характеристика доцента кафедры\\ математической теории
интеллектуальных систем
\\Миронова Андрея Михайловича}
\end{center}

А.М. Миронов с отличием окончил механико-математический факультет МГУ в 1989 году и
поступил в аспирантуру. В 1992 году он защитил диссертацию на соискание ученой степени кандидата физико-математических наук по теме “Морфизмы реакции для автоматов в категориях”.

А.М. Миронов работает на кафедре МаТИС с 2016 года, в настоящий момент занимает должность доцента. За это время им было опубликовано 53 статьи (из них 9 опубликованы
за последние 5 лет). За последние годы А.М. Миронов выступал
с докладами на конференциях “Интеллектуальные системы и компьютерные науки”, всероссийской научной конференции "Математические основы информатики и информационно-телекоммуникационных систем", "10th Computer Science On-line Conference", 
 и др. 

А.М. Миронов ведет активную педагогическую работу. Под его руководством защищено 
32 дипломные работы на механико-математическом факультете МГУ. 

Научная работа А.М. Миронова связана с математическими моделями в компьютерной безопасности, верификацией программ и математической теории обучения с подкреплением. Им получены результаты по 
разработке новой модели криптографических протоколов и основанных на этой модели методов верификации криптографических протоколов, верификации параллельных программ и распределенных алгоритмов, 
моделированию процессов  машинного обучения.

Считаем, что А.М. Миронов достоин быть научным руководителем аспирантов.

\vspace{10mm}

\hspace{80mm}1 сентября 2025 года

\vspace{10mm}

Заведующий кафедрой \\математической
теории интеллектуальных систем,\\
д.ф.-м.н., профессор Э.Э. Гасанов

\end{document}